\documentclass[11pt]{article}
\usepackage{epsfig} 
\usepackage{graphics}
\newcommand{\sect}[1]{ \section{#1} \setcounter{equation}{0} }

\newcommand{\Nf}{N_{\!f}}

\newcommand{\bare}{\mbox{\footnotesize{o}}}

\newcommand{\MSbar}{\overline{\mbox{MS}}}
\newcommand{\MSbars}{\overline{\mbox{\footnotesize{MS}}}}
\newcommand{\MSbarss}{\overline{\mbox{\scriptsize{MS}}}}
\newcommand{\mMOM}{\mbox{mMOM}}
\newcommand{\mMOMs}{\mbox{\footnotesize{mMOM}}}
\newcommand{\RI}{\mbox{RI${}^\prime$}}

\begin{document}

\title{Five loop minimal MOM scheme field and quark mass anomalous dimensions 
in QCD}

\author{J.A. Gracey \& R.H. Mason, \\ Theoretical Physics Division, \\ 
Department of Mathematical Sciences, \\ University of Liverpool, \\ P.O. Box 
147, \\ Liverpool, \\ L69 3BX, \\ United Kingdom.} 
\date{}

\maketitle 

\vspace{5cm} 
\noindent 
{\bf Abstract.} We determine the anomalous dimensions of the gluon, 
Faddeev-Popov ghost and quark in the minimal MOM scheme to five loops for a
general colour group when Quantum Chromodynamics is fixed in a linear covariant
gauge. The quark mass anomalous dimension is also constructed in the same  
scheme.

\vspace{-16.0cm}
\hspace{13.2cm}
{\bf LTH 1319}

\newpage 

\sect{Introduction.}

High order loop calculations in perturbative quantum field theory are carried
out with respect to a renormalization scheme. Invariably the main scheme of
choice is the modified minimal subtraction ($\MSbar$) scheme, \cite{1,2}. It is
defined by the prescription that at the subtraction point of a divergent 
$n$-point function in a renormalizable theory the singularities of the Laurent 
series in the regularization parameter are absorbed into the renormalization 
constant for that Green's function, \cite{1}. In addition a specific finite 
part, which is $4\pi e^{-\gamma}$ where $\gamma$ is the Euler-Mascheroni 
constant, is also removed, \cite{2}. The main benefit of the $\MSbar$ scheme is
that high order loop calculations can be pushed to extremely high order 
analytically. Several impressive examples that illustrate this in recent years 
are the {\em five} loop $\beta$-function of Quantum Chromodynamics (QCD), 
\cite{3,4,5,6,7}, and the renormalization of scalar $\phi^4$ theory at
{\em seven} loops, \cite{8}. While this level of precision for QCD has been 
crucial for phenomenology studies, the $\MSbar$ scheme is in one sense a 
universal scheme. This is because in that scheme the choice for the momentum 
configuration of the divergent $n$-point function where the renormalization 
constants are defined is virtually free. The only places where one has to be 
careful are configurations which are termed exceptional where techniques such 
as infrared rearrangement, \cite{9,10,11,12}, have to be employed. In other 
words the $\MSbar$ scheme is not tied to any kinematic property of the 
$n$-point function. By contrast kinematic schemes are connected to 
non-exceptional momentum configurations and information concerning specific 
properties of the subtraction point are reflected in the structure of the 
renormalization constants. A well-known set of such schemes are those provided 
by Celmaster and Gonsalves in \cite{13,14} for the renormalization of QCD. In 
the three different momentum subtraction (MOM) schemes, each based on one of 
the $3$-point vertices in the QCD Lagrangian, the respective vertex functions 
are evaluated at the fully symmetric point. This is where the square of each of
the external momenta are all equal to each other. Then the renormalization 
prescription is that at this symmetric configuration the renormalization 
constant is chosen so that there are no $O(a)$ corrections where 
$a$~$=$~$g^2/(16\pi^2)$ and $g$ is the gauge coupling constant. While this 
scheme has a phenomenological origin other schemes have been introduced to 
accommodate certain issues. 

One such scheme is the minimal momentum subtraction ($\mMOM$) scheme that was
introduced in \cite{15} to exploit and extend a particular fundamental 
property of QCD that was originally observed by Taylor in \cite{16}. More
specifically it was proved in \cite{16} that the gluon-ghost vertex function is
finite to all orders in the Landau gauge. One consequence is that the 
$\MSbar$ QCD $\beta$-function can therefore be deduced from the Landau gauge 
values of the gluon and ghost anomalous dimensions, \cite{17}. However to ease 
the numerical and financial aspects of making measurements of lattice 
regularized quantities in QCD the $\mMOM$ scheme was developed in such a way 
that the non-renormalization of the gluon-ghost vertex was maintained for an 
arbitrary linear covariant gauge, \cite{15}. Although initially defined for 
lattice regularization of QCD it has a continuum spacetime analogue which was 
given in \cite{15}. In particular the $\mMOM$ $\beta$-function was computed to 
four loops, \cite{15}, with the field anomalous dimensions together with the 
quark mass anomalous dimension following later in \cite{18}. These 
renormalization group functions were required for studying the conformal window
properties of QCD and the associated critical exponents at the Banks-Zaks fixed
point that was discovered in \cite{19,20}. One property of a fixed point of the
$\beta$-function is that the critical exponents, derived from evaluating the 
anomalous dimensions at the fixed point, are renormalization group invariants. 
In other words they are scheme independent. Therefore the four loop $\mMOM$ 
anomalous dimensions were needed to study the convergence of critical exponent 
estimates \cite{21,22}. Given the continued interest in such conformal window 
studies for gauge theories, \cite{21,22,23,24}, and theories beyond the 
Standard Model coupled with the extension of QCD renormalization group 
functions to five loops, the aim of this article is to provide the $\mMOM$ 
field and quark mass anomalous dimensions to the same order. In \cite{25} only 
the five loop $\mMOM$ $\beta$-function was presented and then used to determine
the $R$ ratio in that scheme to a new loop order. Other phenomenological 
applications of the $\mMOM$ scheme were discussed in \cite{26,27}. However, 
using data provided in \cite{25} we have been able to determine the $\mMOM$ 
field and quark mass renormalization constants to four loops. Knowledge of 
these will then allow us to deduce their {\em five} loop $\mMOM$ anomalous 
dimensions from a particular property of the renormalization group. In addition
the anomalous dimensions are also needed for a parallel study of the fixed 
points of QCD, including the Banks-Zaks one, in $\mMOM$, \cite{28}. That 
article examines the fixed point structure in a variety schemes, such as the 
MOM ones of \cite{13,14}, as well as QCD fixed in both linear and nonlinear 
covariant gauges. The $\mMOM$ aspect of that work relies importantly on the 
separate five loop results provided here.

The article is organized as follows. We review the basics of the $\mMOM$ scheme
in Section $2$ as well as outlining the renormalization group formalism we use 
to extract the five loop $\mMOM$ renormalization group functions. The full 
results are recorded in Section $3$ and concluding remarks are provided in 
Section $4$.

\sect{Background.}

The defining criterion for the $\mMOM$ scheme is that the non-renormalization
of the ghost-gluon vertex in the Landau gauge observed by Taylor in \cite{16}
is preserved for a non-zero linear covariant gauge parameter. In practical 
terms this translates into a condition between the relevant renormalization
constants associated with that vertex which is, \cite{15},
\begin{equation}
Z_g^{\MSbarss} \sqrt{Z_A^{\MSbarss}} Z_c^{\MSbarss} ~=~
Z_g^{\mMOMs} \sqrt{Z_A^{\mMOMs}} Z_c^{\mMOMs}
\label{mMOMdef}
\end{equation}
where $Z_g$, $Z_A$ and $Z_c$ are the respective renormalization constants for
the gauge coupling constant $g$ together with the gluon and the Faddeev-Popov 
ghost fields. Each is labelled by the scheme in which the basic variables $a$ 
and the covariant gauge parameter $\alpha$ are in. The $\MSbar$ renormalization
constants have been determined to five loops 
\cite{3,4,5,6,7,17,29,30,31,32,33,34,35,36,37,38} over many years. For the 
$\mMOM$ scheme the gluon and ghost renormalization constants are defined in the
same way as their counterparts in the MOM scheme of \cite{13,14}. As one 
determines the $\mMOM$ field renormalization constants order by order in the 
loop expansion then $Z_g^{\mMOMs}$ is deduced iteratively via (\ref{mMOMdef}). 
This procedure was applied in \cite{15} to determine the four loop $\mMOM$ 
$\beta$-function with the anomalous dimensions of the gluon, ghost and quark as
well as that of the quark mass provided in \cite{18} to the same loop order. In
practical terms to find the four loop $\mMOM$ results the $2$-point functions 
of the fields were computed to three loops using the {\sc Mincer} package, 
\cite{39,40}. At a technical level the $2$-point functions were computed as a 
Laurent expansion in the dimensional regularization parameter $\epsilon$ where 
$d$~$=$~$4$~$-$~$2\epsilon$. Such an expansion is necessary to ensure the 
correct finite part emerges as one applies the $\mMOM$ renormalization 
conditions to find the renormalization constants. Once the three loop $\mMOM$ 
renormalization constants have been determined then the four loop anomalous 
dimensions can be deduced from a property of the renormalization group 
equation. 

To illustrate this we recall the basic relation of the field renormalization
constants to their associated anomalous dimension in the specific scheme that 
they have been determined in, such as either the $\MSbar$ or $\mMOM$ schemes, 
is
\begin{eqnarray}
\gamma_\phi(a,\alpha) &=&
\beta(a,\alpha) \frac{\partial \ln Z_\phi}{\partial a} ~+~
\alpha \gamma_\alpha(a,\alpha)
\frac{\partial \ln Z_\phi}{\partial \alpha}
\label{anomdimdef}
\end{eqnarray}
where $\phi$~$\in$~$\{A,c,\psi\}$ noting that
\begin{eqnarray}
\gamma_\alpha(a,\alpha) &=&
\left[ \beta(a,\alpha) \frac{\partial \ln Z_\alpha}{\partial a} ~-~
\gamma_A(a,\alpha) \right]
\left[ 1 ~-~ \alpha \frac{\partial \ln Z_\alpha}{\partial \alpha}
\right]^{-1}
\label{gamalphadef} 
\end{eqnarray}
for the renormalization of the gauge parameter in general. In a {\it linear}
covariant gauge fixing $Z_\alpha$ is unity whence the latter relation for the
anomalous dimension of $\alpha$ is trivially related to $\gamma_A(a,\alpha)$.
Our convention for the relation between the renormalized and bare, denoted by 
${}_{\bare}$, parameter is
\begin{equation}
\alpha_{\bare} ~=~ Z^{-1}_\alpha Z_A \, \alpha ~.
\end{equation}
Examining (\ref{gamalphadef}) it might appear that there is a singularity when 
the denominator vanishes. This can only occur if 
$Z_\alpha$~$=$~$\lambda \alpha$ where $\lambda$ is an arbitrary non-zero 
constant. However as this is independent of the coupling constant and does not 
commence with unity in a perturbative expansion then the denominator of
(\ref{gamalphadef}) will never vanish. We have taken a general position and 
included $\alpha$ as an argument of the $\beta$-function. In the $\MSbar$ 
scheme the $\beta$-function is independent of the covariant gauge parameter, 
\cite{1}. With (\ref{anomdimdef}) and defining the conversion functions for the
two schemes of interest by 
\begin{equation}
C_\phi(a,\alpha) ~=~ \frac{Z^{\mMOMs}_\phi}{Z^{\MSbarss}_\phi}
\label{convdef}
\end{equation}
where the variables of the conversion functions will always be in the $\MSbar$
scheme, then it is straightforward to relate the anomalous dimensions in one 
scheme with those in the other following a similar approach given in \cite{41}.
This leads to
\begin{eqnarray}
\gamma^{\mMOMs}_\phi \left( a_{\mMOMs}, \alpha_{\mMOMs} \right) &=& \left[
\gamma^{\MSbarss}_\phi \left( a_{\MSbarss}, \alpha_{\MSbarss} \right) +
\beta^{\MSbarss} \left( a_{\MSbarss} \right)
\frac{\partial ~}{\partial a_{\MSbarss}}
\ln C_\phi \left( a_{\MSbarss}, \alpha_{\MSbarss} \right) \right. \nonumber \\
&& \left. \,+\, \alpha_{\MSbarss} \, 
\gamma^{\MSbarss}_\alpha \left( a_{\MSbarss}, \alpha_{\MSbarss} \right) 
\frac{\partial ~}{\partial \alpha_{\MSbarss}}
\ln C_\phi \left( a_{\MSbarss}, \alpha_{\MSbarss} \right)
\right]_{\stackrel{\MSbars \to}{\mMOMs}} ~~~~~~
\label{gammamap}
\end{eqnarray}
where we label the variables and functions by their scheme. The final stage
requires one to map the $\MSbar$ variables on the right side to their $\mMOM$ 
partners to ensure the expression is a function of the correct variables. This 
is achieved by recalling that the relations between the coupling constant and 
gauge parameter in each scheme are given by
\begin{equation}
g_{\mMOMs} ~=~ \frac{Z_g^{\MSbarss}}{Z_g^{\mMOMs}} \, g_{\MSbarss} ~~~,~~~
\alpha_{\mMOMs} ~=~ \frac{Z_A^{\mMOMs}}{Z_A^{\MSbarss}} \alpha_{\MSbarss} ~.
\end{equation}
The explicit expressions can be deduced once the respective renormalization
constants are available at the required order. The key property of
(\ref{anomdimdef}) that allows us to determine the five loop $\mMOM$ anomalous 
dimensions is that the $\mMOM$ renormalization constants are only needed to 
{\it four} loops which can be seen by examining the $a$ dependence of each term
in (\ref{gammamap}). More specifically the first term on the right side of
(\ref{gammamap}) is available to $O(a_{\MSbarss}^5)$ as are
$\beta^{\MSbarss} \left( a_{\MSbarss} \right)$ and
$\gamma^{\MSbarss}_\alpha \left( a_{\MSbarss}, \alpha_{\MSbarss} \right)$.
However examining (\ref{convdef}) and using the explicit expressions for
$Z^{\mMOMs}_\phi$, that we will deduce at $O(a_{\MSbarss}^4)$ momentarily, 
means that
$\frac{\partial ~}{\partial a_{\MSbarss}}
\ln C_\phi \left( a_{\MSbarss}, \alpha_{\MSbarss} \right)$
and $\frac{\partial ~}{\partial \alpha_{\MSbarss}}
\ln C_\phi \left( a_{\MSbarss}, \alpha_{\MSbarss} \right)$
will be available to $O(a_{\MSbarss}^3)$ and $O(a_{\MSbarss}^4)$ respectively. 
As the leading terms of $\beta^{\MSbarss} \left( a_{\MSbarss} \right)$ and
$\gamma^{\MSbarss}_\alpha \left( a_{\MSbarss}, \alpha_{\MSbarss} \right)$ are 
$O(a_{\MSbarss}^2)$ and $O(a_{\MSbarss})$ respectively then when the 
combinations of the final two terms of (\ref{gammamap}) are compiled the 
$O(a_{\MSbarss}^5)$ coefficient of each anomalous dimension is known. Therefore
all that remains in this exercise is to determine the four loop $\mMOM$ 
renormalization constants.

This can be achieved relatively straightforwardly from the data presented in 
\cite{25}. In \cite{25} the expressions for the $2$-point functions as well as 
the $3$-point vertices, where there was a nullified external momentum on one of
the fields, were presented as a function of the bare parameters. Moreover the 
Laurent expansion in powers of $\epsilon$ was given to four loops. This was 
determined via the {\sc Forcer} algorithm, \cite{42,43}, and formed the basis 
of calculating the five loop $\MSbar$ renormalization group functions of QCD. 
Since the expressions are available in terms of bare parameters there is 
sufficient data for the four loop $\mMOM$ renormalization constants to be found
through the scheme prescription described earlier. In extracting these and 
converting them to the renormalization group functions in the $\mMOM$ scheme 
using {\sc Form}, \cite{44,45}, we have verified the four loop expressions 
given in \cite{18}. We have carried out the same procedure for the quark mass 
anomalous dimension. In this instance we made use of the four loop 
renormalization of the quark mass operator provided in \cite{46}. The operator 
was renormalized in that article in the $\RI$ scheme which is a scheme 
underpinning the matching of lattice field theory results to continuum 
perturbation theory. In particular the operator was inserted in a quark 
$2$-point function at zero momentum. While the $\MSbar$ mass dimension has been
known to five loops in the $\MSbar$ scheme for several years now 
\cite{36,37,38,47,48,49,50,51}, the finite part of the Green's function was not
available at four loops in terms of bare parameters to high enough order in the
$\epsilon$ expansion to implement renormalization in schemes whose 
prescriptions require finite parts to be absorbed into the renormalization 
constants. As this is the case for our $\mMOM$ analysis we have used the 
results of \cite{46}, where the quark mass operator was inserted at zero 
momentum in a quark $2$-point function, to determine the four loop quark mass 
renormalization constant in the $\mMOM$ scheme. 

\sect{Results.}

Having established the formalism and strategy to extract the five loop $\mMOM$ 
renormalization group functions we devote this section to recording the various
expressions. In evaluating Green's functions to higher loop order it is 
well-known that the number of Feynman graphs increases. As a consequence the 
resulting expressions become larger in QCD due to additional colour group
factors being introduced through new graph topologies such as diagrams that
involve so called light-by-light structures. Therefore we illustrate the 
essence of our results by recording the relevant expressions for the $SU(3)$ 
colour group in the Landau gauge. The full gauge dependence for $SU(3)$ is 
given in the Appendix while we provide the expressions for an arbitrary colour 
group in a data file associated with the arXiv version of this paper. First, we 
record the four conversion functions are
\begin{eqnarray}
\left. C_A(a,0) \right|^{SU(3)} &=& 1
+ \left[
\frac{97}{12}
- \frac{10}{9} \Nf
\right] a
+ \left[
\frac{83105}{288}
- \frac{11299}{216} \Nf
- \frac{4}{3} \zeta_3 \Nf
+ \frac{100}{81} \Nf^2
- 27 \zeta_3
\right] a^2
\nonumber \\
&&
+ \left[
\frac{164395}{972} \Nf^2
- \frac{8228977}{2592} \Nf
- \frac{63225}{64} \zeta_5
- \frac{17433}{8} \zeta_3
- \frac{1000}{729} \Nf^3
- \frac{243}{32} \zeta_4
\right. \nonumber \\
&& \left. ~~~
+ \frac{16}{9} \zeta_3 \Nf^2
+ \frac{33}{2} \zeta_4 \Nf
+ \frac{2320}{9} \zeta_5 \Nf
+ \frac{25915}{108} \zeta_3 \Nf
+ \frac{44961125}{3456}
\right] a^3
\nonumber \\
&&
+ \left[
\frac{62302764631}{82944}
- \frac{1758762815}{7776} \Nf
- \frac{324121925}{2048} \zeta_3
- \frac{262747689}{1024} \zeta_5
\right. \nonumber \\
&& \left. ~~~
- \frac{25666081}{864} \zeta_7 \Nf
- \frac{15059695}{34992} \Nf^3
- \frac{3161781}{2048} \zeta_4
- \frac{515843}{162} \zeta_5 \Nf^2
\right. \nonumber \\
&& \left. ~~~
- \frac{362555}{648} \zeta_3 \Nf^2
- \frac{174483}{32} \zeta_3^2
- \frac{16775}{16} \zeta_6 \Nf
- \frac{1943}{216} \zeta_4 \Nf^2
- \frac{1107}{8} \zeta_3^2 \Nf
\right. \nonumber \\
&& \left. ~~~
- \frac{229}{27} \zeta_3 \Nf^3
+ \frac{400}{9} \zeta_3^2 \Nf^2
+ \frac{880}{27} \zeta_5 \Nf^3
+ \frac{10000}{6561} \Nf^4
+ \frac{204525}{32} \zeta_6
\right. \nonumber \\
&& \left. ~~~
+ \frac{469333}{576} \zeta_4 \Nf
+ \frac{27270475}{864} \zeta_3 \Nf
+ \frac{60587905}{864} \zeta_5 \Nf
+ \frac{190787741}{10368} \Nf^2
\right. \nonumber \\
&& \left. ~~~
+ \frac{277127487}{2048} \zeta_7
- \zeta_4 \Nf^3
\right] a^4 ~+~ O(a^5)
\end{eqnarray}
\begin{eqnarray}
\left. C_c(a,0) \right|^{SU(3)} &=& 1 + 3 a
+ \left[
\frac{5829}{64}
- \frac{135}{16} \zeta_3
- \frac{95}{16} \Nf
\right] a^2
\nonumber \\
&&
+ \left[
\frac{5161}{648} \Nf^2
- \frac{198001}{432} \Nf
- \frac{40449}{64} \zeta_3
- \frac{1755}{32} \zeta_5
- \frac{33}{4} \zeta_4 \Nf
+ \frac{2}{3} \zeta_3 \Nf^2
\right. \nonumber \\
&& \left. ~~~
+ \frac{59}{2} \zeta_3 \Nf
+ \frac{243}{64} \zeta_4
+ \frac{1082353}{288}
\right] a^3
\nonumber \\
&&
+ \left[
\frac{15567976783}{73728}
- \frac{151911987}{4096} \zeta_3
- \frac{79190001}{2048} \zeta_5
- \frac{39621021}{1024} \Nf
\right. \nonumber \\
&& \left. ~~~
- \frac{204525}{64} \zeta_6
- \frac{150979}{7776} \Nf^3
- \frac{52025}{128} \zeta_4 \Nf
- \frac{11907}{16} \zeta_7 \Nf
- \frac{699}{8} \zeta_5 \Nf^2
\right. \nonumber \\
&& \left. ~~~
- \frac{425}{48} \zeta_3 \Nf^2
- \frac{5}{18} \zeta_3 \Nf^3
+ \frac{1}{2} \zeta_4 \Nf^3
+ \frac{165}{16} \zeta_4 \Nf^2
+ \frac{337}{4} \zeta_3^2 \Nf
+ \frac{16775}{32} \zeta_6 \Nf
\right. \nonumber \\
&& \left. ~~~
+ \frac{250085}{64} \zeta_5 \Nf
+ \frac{1149471}{512} \zeta_3^2
+ \frac{2625583}{768} \zeta_3 \Nf
+ \frac{5819877}{4096} \zeta_4
+ \frac{8093153}{4608} \Nf^2
\right. \nonumber \\
&& \left. ~~~
+ \frac{100880073}{8192} \zeta_7
\right] a^4 ~+~ O(a^5)
\end{eqnarray}
and
\begin{eqnarray}
\left. C_\psi(a,0) \right|^{SU(3)} &=& 1 
+ \left[
\frac{7}{3} \Nf
+ 12 \zeta_3
- \frac{359}{9}
\right] a^2
\nonumber \\
&&
+ \left[
\frac{24722}{81} \Nf
- \frac{439543}{162}
- \frac{1570}{243} \Nf^2
- \frac{1165}{3} \zeta_5
- \frac{440}{9} \zeta_3 \Nf
+ \frac{79}{4} \zeta_4
\right. \nonumber \\
&& \left. ~~~
+ \frac{8009}{6} \zeta_3
\right] a^3
\nonumber \\
&&
+ \left[
\frac{21391}{1458} \Nf^3
- \frac{356864009}{5184} \zeta_5
- \frac{146722043}{864}
- \frac{29889697}{5184} \zeta_3^2
- \frac{1294381}{108} \zeta_3 \Nf
\right. \nonumber \\
&& \left. ~~~
- \frac{1276817}{972} \Nf^2
- \frac{440}{9} \zeta_5 \Nf^2
- \frac{20}{3} \zeta_4 \Nf^2
+ \frac{8}{27} \zeta_3 \Nf^3
+ \frac{100}{3} \zeta_6 \Nf
\right. \nonumber \\
&& \left. ~~~
+ \frac{2291}{72} \zeta_4 \Nf
+ \frac{5704}{27} \zeta_3 \Nf^2
+ \frac{565939}{864} \zeta_4
+ \frac{1673051}{324} \zeta_5 \Nf
+ \frac{3807625}{10368} \zeta_6
\right. \nonumber \\
&& \left. ~~~
+ \frac{6747755}{288} \zeta_7
+ \frac{55476671}{1944} \Nf
+ \frac{317781451}{2592} \zeta_3
- 1029 \zeta_7 \Nf
- 24 \zeta_3^2 \Nf
\right] a^4 
\nonumber \\
&& +~ O(a^5)
\end{eqnarray}
for the fields and
\begin{eqnarray}
\left. C_m(a,0) \right|^{SU(3)} &=& 1
- \frac{16}{3} a
+ \left[
\frac{83}{9} \Nf
+ \frac{152}{3} \zeta_3
- \frac{3779}{18}
\right] a^2
\nonumber \\
&&
+ \left[
\frac{217390}{243} \Nf
- \frac{3115807}{324}
- \frac{7514}{729} \Nf^2
- \frac{4720}{27} \zeta_3 \Nf
- \frac{2960}{9} \zeta_5
- \frac{32}{27} \zeta_3 \Nf^2
\right. \nonumber \\
&& \left. ~~~
+ \frac{80}{3} \zeta_4 \Nf
+ \frac{195809}{54} \zeta_3
\right] a^3
\nonumber \\
&&
+ \left[
\frac{40}{81} \zeta_3 \Nf^3
- \frac{744609145}{1296}
- \frac{52383125}{17496} \Nf^2
- \frac{3837631}{1728} \zeta_7
- \frac{2017309}{81} \zeta_3 \Nf
\right. \nonumber \\
&& \left. ~~~
- \frac{843077}{54} \zeta_3^2
- \frac{359855}{81} \zeta_5 \Nf
- \frac{16960}{9} \zeta_4
- \frac{11500}{9} \zeta_6 \Nf
- \frac{8776}{27} \zeta_3^2 \Nf
\right. \nonumber \\
&& \left. ~~~
- \frac{343}{2} \zeta_7 \Nf
- \frac{100}{3} \zeta_4 \Nf^2
- \frac{8}{9} \zeta_4 \Nf^3
+ \frac{560}{3} \zeta_5 \Nf^2
+ \frac{11542}{9} \zeta_4 \Nf
\right. \nonumber \\
&& \left. ~~~
+ \frac{33964}{81} \zeta_3 \Nf^2
+ \frac{96979}{4374} \Nf^3
+ \frac{9369745}{432} \zeta_5
+ \frac{86284171}{324} \zeta_3
+ \frac{247516535}{2916} \Nf
\right. \nonumber \\
&& \left. ~~~
+ 5500 \zeta_6
\right] a^4 ~+~ O(a^5)
\end{eqnarray}
for the quark mass renormalization where $\zeta_n$ is the Riemann zeta function
and $\Nf$ is the number of quark flavours. One minor check on the derivation of
these expressions is that they are finite with respect to $\epsilon$ given that
they derive from the ratio of four loop renormalization constants.

The conversion functions coupled with the five loop $\MSbar$ anomalous
dimensions derived in \cite{5,7,36,37,38} for an arbitrary colour group mean 
that we can extract the corresponding $\mMOM$ expressions to the same loop 
order via (\ref{gammamap}). Again restricting to the case of Landau gauge and 
the $SU(3)$ group we arrive at our main results which are
\begin{eqnarray}
\left. \gamma_A^{\mMOMs}(a,0) \right|^{SU(3)} &=&
\left[
\frac{2}{3} \Nf
- \frac{13}{2}
\right] a
+ \left[
 \frac{67}{6} \Nf
- \frac{255}{4}
\right] a^2
\nonumber \\
&&
+ \left[
\frac{11227}{24} \Nf
- \frac{8637}{4}
- \frac{719}{54} \Nf^2
- \frac{229}{12} \zeta_3 \Nf
- \frac{8}{9} \zeta_3 \Nf^2
+ 324 \zeta_3
\right] a^3
\nonumber \\
&&
+ \left[
\frac{5549393}{192} \Nf
- \frac{27189875}{256}
- \frac{1118977}{648} \Nf^2
- \frac{921265}{144} \zeta_5 \Nf
\right. \nonumber \\
&& \left. ~~~
- \frac{889231}{144} \zeta_3 \Nf
- \frac{16}{9} \zeta_3 \Nf^3
+ \frac{665}{27} \Nf^3
+ \frac{5143}{27} \zeta_3 \Nf^2
+ \frac{9280}{27} \zeta_5 \Nf^2
\right. \nonumber \\
&& \left. ~~~
+ \frac{1950705}{128} \zeta_5
+ \frac{7740879}{256} \zeta_3
\right] a^4
\nonumber \\
&& 
+ \left[
\frac{1760}{27} \zeta_5 \Nf^4
- \frac{71363464263}{16384} \zeta_7
- \frac{16520894997}{2048}
- \frac{16359945025}{6912} \zeta_5 \Nf
\right. \nonumber \\
&& \left. ~~~
- \frac{8539017539}{20736} \zeta_3 \Nf
+ \frac{3150668061}{2048} \zeta_3
+ \frac{23633674547}{18432} \zeta_7 \Nf
\right. \nonumber \\
&& \left. ~~~
+ \frac{29623505625}{4096} \zeta_5
+ \frac{327291152977}{124416} \Nf
- \frac{613783375}{2592} \Nf^2
\right. \nonumber \\
&& \left. ~~~
- \frac{25237429}{432} \zeta_7 \Nf^2
- \frac{979887}{64} \zeta_3^2 \Nf
- \frac{292855}{54} \zeta_5 \Nf^3
- \frac{51647}{36} \zeta_3^2 \Nf^2
\right. \nonumber \\
&& \left. ~~~
- \frac{1861}{27} \Nf^4
- \frac{304}{27} \zeta_3 \Nf^4
+ \frac{2240}{27} \zeta_3^2 \Nf^3
+ \frac{129545}{162} \zeta_3 \Nf^3
\right. \nonumber \\
&& \left. ~~~
+ \frac{2420705}{324} \Nf^3
+ \frac{13148441}{2592} \zeta_3 \Nf^2
+ \frac{246191965}{1296} \zeta_5 \Nf^2
\right. \nonumber \\
&& \left. ~~~
+ \frac{671260095}{2048} \zeta_3^2
\right] a^5 ~+~ O(a^6)
\end{eqnarray}
\begin{eqnarray}
\left. \gamma_c^{\mMOMs}(a,0) \right|^{SU(3)} &=&
-~ \frac{9}{4} a
+ \left[
\frac{3}{4} \Nf
- \frac{153}{8}
\right] a^2
\nonumber \\
&&
+ \left[
\frac{1401}{16} \Nf
- \frac{11691}{16}
- \frac{5}{2} \Nf^2
+ \frac{9}{4} \zeta_3 \Nf
+ \frac{1269}{16} \zeta_3
\right] a^3
\nonumber \\
&&
+ \left[
\frac{5}{2} \Nf^3
- \frac{16985133}{512}
- \frac{16059}{32} \zeta_3 \Nf
- \frac{13115}{48} \Nf^2
- \frac{5895}{16} \zeta_5 \Nf
\right. \nonumber \\
&& \left. ~~~
+ \frac{739053}{128} \Nf
+ \frac{1140075}{256} \zeta_5
+ \frac{2264193}{512} \zeta_3
+ 26 \zeta_3 \Nf^2
\right] a^4
\nonumber \\
&&
+ \left[
\frac{9758887695}{8192} \zeta_5
- \frac{15114361941}{32768} \zeta_7
- \frac{9464865363}{4096}
- \frac{34433025}{256} \zeta_5 \Nf
\right. \nonumber \\
&& \left. ~~~
- \frac{12230645}{288} \Nf^2
- \frac{86231}{2} \zeta_3 \Nf
- \frac{3969}{2} \zeta_7 \Nf^2
- \frac{219}{2} \zeta_3^2 \Nf^2
\right. \nonumber \\
&& \left. ~~~
+ \frac{15353}{12} \Nf^3
+ \frac{477395}{192} \zeta_3 \Nf^2
+ \frac{550645}{96} \zeta_5 \Nf^2
+ \frac{814149}{128} \zeta_3^2 \Nf
\right. \nonumber \\
&& \left. ~~~
+ \frac{34354859}{1024} \zeta_7 \Nf
+ \frac{1106092719}{4096} \zeta_3
+ \frac{1670942731}{3072} \Nf
\right. \nonumber \\
&& \left. ~~~
- \frac{424982943}{4096} \zeta_3^2
- 65 \zeta_5 \Nf^3
- 14 \Nf^4
+ \zeta_3 \Nf^3
\right] a^5
~+~ O(a^6)
\end{eqnarray}
\begin{eqnarray}
\left. \gamma_\psi^{\mMOMs}(a,0) \right|^{SU(3)} &=&
\left[
\frac{67}{3}
- \frac{4}{3} \Nf
\right] a^2
+ \left[
\frac{29675}{36}
- \frac{706}{9} \Nf
- \frac{607}{2} \zeta_3
+ \frac{8}{9} \Nf^2
+ 16 \zeta_3 \Nf
\right] a^3
\nonumber \\
&&
+ \left[
\frac{31003343}{648}
- \frac{21683117}{648} \zeta_3
- \frac{2393555}{324} \Nf
- \frac{272}{9} \zeta_3 \Nf^2
- \frac{40}{9} \Nf^3
\right. \nonumber \\
&& \left. ~~~
+ \frac{2861}{9} \Nf^2
+ \frac{74440}{27} \zeta_3 \Nf
+ \frac{15846715}{1296} \zeta_5
- 830 \zeta_5 \Nf
\right] a^4
\nonumber \\
&&
+ \left[
\frac{2313514793}{10368} \zeta_3^2
- \frac{94958116621}{31104} \zeta_3
- \frac{26588447977}{27648} \zeta_7
\right. \nonumber \\
&& \left. ~~~
+ \frac{14723323093}{5184}
+ \frac{18607183745}{7776} \zeta_5
- \frac{667846415}{1944} \zeta_5 \Nf
\right. \nonumber \\
&& \left. ~~~
- \frac{251804567}{432} \Nf
- \frac{4726621}{243} \zeta_3 \Nf^2
- \frac{596849}{54} \zeta_3^2 \Nf
- \frac{80606}{81} \Nf^3
\right. \nonumber \\
&& \left. ~~~
- \frac{6811}{2} \zeta_7 \Nf^2
- \frac{3520}{27} \zeta_5 \Nf^3
- \frac{128}{9} \zeta_3^2 \Nf^2
+ \frac{160}{27} \Nf^4
+ \frac{24064}{81} \zeta_3 \Nf^3
\right. \nonumber \\
&& \left. ~~~
+ \frac{1063237}{27} \Nf^2
+ \frac{3085750}{243} \zeta_5 \Nf^2
+ \frac{4429579}{36} \zeta_7 \Nf
\right. \nonumber \\
&& \left. ~~~
+ \frac{1748707267}{3888} \zeta_3 \Nf
\right] a^5 ~+~ O(a^6)
\end{eqnarray}
and
\begin{eqnarray}
\left. \gamma_m^{\mMOMs}(a,0) \right|^{SU(3)} &=&
-~ 4 a
+ \left[
\frac{4}{3} \Nf
- \frac{209}{3}
\right] a^2
\nonumber \\
&&
+ \left[
\frac{5635}{6} \zeta_3
- \frac{95383}{36}
- \frac{176}{9} \zeta_3 \Nf
- \frac{8}{3} \Nf^2
+ \frac{4742}{27} \Nf
\right] a^3
\nonumber \\
&&
+ \left[
\frac{8}{3} \Nf^3
- \frac{182707879}{1296}
- \frac{309295}{48} \zeta_5
- \frac{159817}{27} \zeta_3 \Nf
- \frac{13651}{27} \Nf^2
\right. \nonumber \\
&& \left. ~~~
- \frac{3200}{9} \zeta_5 \Nf
+ \frac{1552}{9} \zeta_3 \Nf^2
+ \frac{5246557}{324} \Nf
+ \frac{15752321}{216} \zeta_3
\right] a^4
\nonumber \\
&&
+ \left[
\frac{3576071485}{27648} \zeta_7
- \frac{75504232175}{7776}
+ \frac{9610932889}{5832} \Nf
+ \frac{17917034005}{31104} \zeta_5
\right. \nonumber \\
&& \left. ~~~
+ \frac{187324052147}{31104} \zeta_3
- \frac{310328447}{432} \zeta_3^2
- \frac{257106335}{324} \zeta_3 \Nf
\right. \nonumber \\
&& \left. ~~~
- \frac{180251015}{1944} \zeta_5 \Nf
- \frac{22459484}{243} \Nf^2
- \frac{4778536}{81} \zeta_7 \Nf
- \frac{60928}{81} \zeta_3^2 \Nf^2
\right. \nonumber \\
&& \left. ~~~
- \frac{28096}{81} \zeta_3 \Nf^3
- \frac{1600}{9} \zeta_5 \Nf^3
- \frac{352}{27} \Nf^4
+ \frac{1372}{3} \zeta_7 \Nf^2
+ \frac{464038}{243} \Nf^3
\right. \nonumber \\
&& \left. ~~~
+ \frac{948548}{27} \zeta_3 \Nf^2
+ \frac{1850845}{243} \zeta_5 \Nf^2
+ \frac{6570181}{162} \zeta_3^2 \Nf
\right] a^5
\nonumber \\
&& +~ O(a^6) 
\end{eqnarray}
where the variables are in the scheme indicated by the label on the
renormalization group function. While we have recorded the Landau gauge 
expressions for the field and quark mass anomalous dimensions an independent 
check on our formalism and results is that we have reproduced the five loop 
$\mMOM$ $\beta$-function for non-zero $\alpha$ and a general colour group. 
Although the Landau gauge result was provided in \cite{25} the full gauge 
dependent result was given in the associated data file of the arXiv version of 
that publication. We have reproduced this precisely for all $\alpha$ which 
means we have used the correct mapping of the $\mMOM$ values of $a$ and 
$\alpha$ to their $\MSbar$ counterparts, and their inverses, in applying the 
equation analogous to that of (\ref{gammamap}) for the $\beta$-function. The 
same data file contains the coupling constant map, which we have independently 
verified, but that for the gauge parameter was not available. For completeness 
we record the $SU(3)$ version for that mapping in the Appendix for 
$\alpha$~$\neq$~$0$ in our conventions. The full colour group expression is
available in the arXiv data file associated with this article.

\sect{Discussion.}

To summarize we have derived the renormalization group functions for the 
gluon, ghost and quark in QCD at five loops in the $\mMOM$ scheme as well as
the quark mass anomalous dimension. With the five loop $\mMOM$ $\beta$-function
of \cite{25} the set of core QCD renormalization group functions in that scheme
is now complete since the running of the linear covariant gauge parameter is
given by $-$~$\gamma^{\mMOMs}_A(a,\alpha)$. Although the $\beta$-function is 
ordinarily of primary importance since it governs the running of the gauge 
coupling constant, with applications to phenomenology \cite{25}, the anomalous 
dimensions are important for conformal window studies, \cite{21,22,23,24}. 
These can now be extended to higher order in relation to the known Banks-Zaks 
infrared fixed point in the $\mMOM$ scheme. While the conformal window has been
studied at length in the case of the $SU(3)$ group due to the connection with 
the strong force of the Standard Model, our results for an arbitrary colour 
group mean that the properties of gauge groups can be explored. Moreover, such 
investigations are not limited to the case where quarks are in the fundamental 
representation since the properties of the conformal window can be explored for
other representations that will have applications for beyond the Standard Model
physics. For instance there have been several lattice studies of some of these 
issues that have relied on the perturbative $\mMOM$ renormalization group 
functions, \cite{52,53}. In the former article the conformal window was 
examined in $SU(2)$ when there are three Majorana fermions in the adjoint 
representation with the aim of understanding the scaling dimensions of bound 
states in the theory. By contrast in \cite{53} the $\mMOM$ scheme results were 
also of importance in exploring ideas concerning lepton compositeness in the 
Standard Model. So five loop $\mMOM$ expressions should assist with improving
the precision of such analyses.

\vspace{1cm}
\noindent
{\bf Acknowledgements.} This work was carried out with the support of the STFC
Consolidated Grant ST/T000988/1 (JAG) and an EPSRC Studentship EP/R513271/1 
(RHM). For the purpose of open access, the authors have applied a Creative 
Commons Attribution (CC-BY) licence to any Author Accepted Manuscript version 
arising. The data representing the main results here are accessible in 
electronic form from the arXiv ancillary directory associated with the article.

\appendix

\sect{Gauge parameter dependent expressions.}

We devote the Appendix to providing the full gauge dependence of the $SU(3)$
anomalous dimensions so that the degree of the polynomial coefficients in 
$\alpha$ at each loop order is manifest. We have 
\begin{eqnarray}
\left. \gamma_A^{\mMOMs}(a,\alpha) \right|^{SU(3)} &=&
\left[
\frac{2}{3} \Nf
+ \frac{3}{2} \alpha
- \frac{13}{2}
\right] a
\nonumber \\
&&
+ \left[
 \frac{67}{6} \Nf
- \frac{255}{4}
- \frac{9}{4} \alpha^3
+ \frac{51}{8} \alpha
+ \frac{51}{8} \alpha^2
- \Nf \alpha
- \Nf \alpha^2
\right] a^2
\nonumber \\
&&
+ \left[
\frac{11227}{24} \Nf
- \frac{8637}{4}
- \frac{1161}{16} \zeta_3 \alpha
- \frac{719}{54} \Nf^2
- \frac{495}{32} \alpha^4
- \frac{237}{8} \Nf \alpha^2
\right. \nonumber \\
&& \left. ~~~
- \frac{229}{12} \zeta_3 \Nf
- \frac{189}{16} \alpha^3
- \frac{139}{8} \Nf \alpha
- \frac{9}{2} \zeta_3 \Nf \alpha
- \frac{9}{4} \Nf \alpha^3
- \frac{8}{9} \zeta_3 \Nf^2
\right. \nonumber \\
&& \left. ~~~
+ \frac{3}{8} \Nf \alpha^4
+ \frac{9}{4} \zeta_3 \Nf \alpha^2
+ \frac{81}{16} \zeta_3 \alpha^3
+ \frac{153}{16} \alpha
+ \frac{5283}{32} \alpha^2
- 54 \zeta_3 \alpha^2
\right. \nonumber \\
&& \left. ~~~
+ 324 \zeta_3
\right] a^3
\nonumber \\
&&
+ \left[
\frac{5549393}{192} \Nf
- \frac{27189875}{256}
- \frac{1815015}{128} \zeta_3 \alpha
- \frac{1118977}{648} \Nf^2
\right. \nonumber \\
&& \left. ~~~
- \frac{921265}{144} \zeta_5 \Nf
- \frac{889231}{144} \zeta_3 \Nf
- \frac{231705}{64} \zeta_3 \alpha^2
- \frac{98423}{64} \Nf \alpha^2
\right. \nonumber \\
&& \left. ~~~
- \frac{65295}{128} \alpha^4
- \frac{50291}{32} \Nf \alpha
- \frac{20727}{256} \alpha^3
- \frac{11745}{128} \alpha^5
- \frac{3579}{32} \Nf \alpha^3
\right. \nonumber \\
&& \left. ~~~
- \frac{1485}{8} \zeta_5 \Nf \alpha
- \frac{945}{32} \zeta_5 \alpha^5
- \frac{495}{64} \zeta_5 \alpha^4
- \frac{315}{8} \zeta_5 \Nf \alpha^2
- \frac{105}{16} \zeta_5 \Nf \alpha^4
\right. \nonumber \\
&& \left. ~~~
- \frac{16}{3} \zeta_3 \Nf^2 \alpha^2
- \frac{16}{9} \zeta_3 \Nf^3
- \frac{15}{8} \zeta_3 \Nf \alpha^4
- \frac{8}{9} \Nf^3 \alpha
+ \frac{9}{4} \Nf \alpha^5
\right. \nonumber \\
&& \left. ~~~
+ \frac{15}{8} \zeta_5 \Nf \alpha^3
+ \frac{27}{4} \zeta_3 \alpha^5
+ \frac{32}{9} \zeta_3 \Nf^3 \alpha
+ \frac{123}{4} \zeta_3 \Nf \alpha^3
+ \frac{417}{32} \Nf \alpha^4
\right. \nonumber \\
&& \left. ~~~
+ \frac{637}{8} \Nf^2 \alpha
+ \frac{665}{27} \Nf^3
+ \frac{1375}{36} \Nf^2 \alpha^2
+ \frac{5143}{27} \zeta_3 \Nf^2
+ \frac{6013}{16} \zeta_3 \Nf \alpha^2
\right. \nonumber \\
&& \left. ~~~
+ \frac{9280}{27} \zeta_5 \Nf^2
+ \frac{18939}{8} \zeta_3 \Nf \alpha
+ \frac{19215}{128} \zeta_3 \alpha^3
+ \frac{23175}{128} \zeta_5 \alpha^3
\right. \nonumber \\
&& \left. ~~~
+ \frac{52821}{256} \zeta_3 \alpha^4
+ \frac{336285}{128} \zeta_5 \alpha^2
+ \frac{753165}{128} \zeta_5 \alpha
+ \frac{1028403}{256} \alpha
\right. \nonumber \\
&& \left. ~~~
+ \frac{1386747}{256} \alpha^2
+ \frac{1950705}{128} \zeta_5
+ \frac{7740879}{256} \zeta_3
- 170 \zeta_3 \Nf^2 \alpha
\right] a^4
\nonumber \\
&& 
+ \left[
\frac{1760}{27} \zeta_5 \Nf^4
- \frac{71363464263}{16384} \zeta_7
- \frac{16520894997}{2048}
\right. \nonumber \\
&& \left. ~~~
- \frac{16359945025}{6912} \zeta_5 \Nf
- \frac{8999200665}{16384} \zeta_7 \alpha
- \frac{8539017539}{20736} \zeta_3 \Nf
\right. \nonumber \\
&& \left. ~~~
+ \frac{2743735509}{4096} \zeta_5 \alpha
+ \frac{3150668061}{2048} \zeta_3
+ \frac{3614875345}{18432} \alpha
\right. \nonumber \\
&& \left. ~~~
+ \frac{23633674547}{18432} \zeta_7 \Nf
+ \frac{29623505625}{4096} \zeta_5
+ \frac{327291152977}{124416} \Nf
\right. \nonumber \\
&& \left. ~~~
- \frac{1470309201}{2048} \zeta_3 \alpha
- \frac{613783375}{2592} \Nf^2
- \frac{498113341}{4608} \Nf \alpha^2
\right. \nonumber \\
&& \left. ~~~
- \frac{453256775}{4608} \Nf \alpha
- \frac{89434611}{2048} \zeta_3^2 \alpha
- \frac{35716707}{2048} \zeta_5 \alpha^3
\right. \nonumber \\
&& \left. ~~~
- \frac{34041915}{2048} \alpha^4
- \frac{32200263}{4096} \zeta_5 \alpha^4
- \frac{32074623}{512} \zeta_7 \alpha^2
\right. \nonumber \\
&& \left. ~~~
- \frac{26500527}{2048} \alpha^3
- \frac{25237429}{432} \zeta_7 \Nf^2
- \frac{24649187}{384} \zeta_5 \Nf \alpha
\right. \nonumber \\
&& \left. ~~~
- \frac{14459067}{16384} \zeta_7 \alpha^4
- \frac{11789955}{64} \zeta_3 \alpha^2
- \frac{11441745}{16384} \zeta_7 \alpha^5
\right. \nonumber \\
&& \left. ~~~
- \frac{5644053}{1024} \zeta_3^2 \alpha^2
- \frac{4574791}{288} \zeta_3 \Nf^2 \alpha
- \frac{3830409}{1024} \zeta_3^2 \alpha^3
- \frac{2526579}{512} \alpha^5
\right. \nonumber \\
&& \left. ~~~
- \frac{1410255}{4096} \zeta_5 \alpha^5
- \frac{1348137}{2048} \zeta_3^2 \alpha^4
- \frac{1058511}{512} \Nf \alpha^3
- \frac{979887}{64} \zeta_3^2 \Nf
\right. \nonumber \\
&& \left. ~~~
- \frac{839241}{256} \zeta_4 \alpha
- \frac{746433}{128} \zeta_3 \Nf \alpha^3
- \frac{520425}{1024} \zeta_5 \alpha^6
- \frac{407025}{512} \zeta_6 \alpha^2
\right. \nonumber \\
&& \left. ~~~
- \frac{318969}{1024} \zeta_7 \Nf \alpha^3
- \frac{292855}{54} \zeta_5 \Nf^3
- \frac{192423}{256} \zeta_3 \Nf \alpha^4
\right. \nonumber \\
&& \left. ~~~
- \frac{186725}{144} \zeta_5 \Nf^2 \alpha^2
- \frac{151875}{256} \zeta_6 \Nf \alpha
- \frac{112077}{2048} \zeta_7 \Nf \alpha^4
\right. \nonumber \\
&& \left. ~~~
- \frac{100115}{144} \zeta_3 \Nf^2 \alpha^2
- \frac{95607}{1024} \zeta_3 \alpha^6
- \frac{76275}{512} \zeta_6 \alpha^4
- \frac{54405}{128} \alpha^6
\right. \nonumber \\
&& \left. ~~~
- \frac{51647}{36} \zeta_3^2 \Nf^2
- \frac{50787}{512} \zeta_4 \alpha^3
- \frac{41993}{243} \Nf^3 \alpha
- \frac{14175}{256} \zeta_6 \Nf \alpha^2
\right. \nonumber \\
&& \left. ~~~
- \frac{10725}{128} \zeta_5 \Nf \alpha^5
- \frac{10125}{256} \zeta_6 \Nf \alpha^3
- \frac{8967}{8} \zeta_7 \Nf^2 \alpha
- \frac{7569}{128} \zeta_3 \Nf \alpha^5
\right. \nonumber \\
&& \left. ~~~
- \frac{4131}{512} \zeta_4 \alpha^5
- \frac{2081}{24} \Nf^2 \alpha^3
- \frac{1861}{27} \Nf^4
- \frac{1549}{18} \Nf^3 \alpha^2
\right. \nonumber \\
&& \left. ~~~
- \frac{1431}{128} \zeta_4 \Nf \alpha^2
- \frac{1131}{16} \zeta_4 \Nf^2 \alpha
- \frac{459}{128} \zeta_4 \Nf \alpha^4
- \frac{304}{27} \zeta_3 \Nf^4
\right. \nonumber \\
&& \left. ~~~
- \frac{260}{3} \zeta_5 \Nf^3 \alpha
- \frac{32}{3} \zeta_3 \Nf^3 \alpha^3
- \frac{21}{4} \zeta_3^2 \Nf^2 \alpha^2
- \frac{15}{2} \zeta_5 \Nf^2 \alpha^3
+ \frac{5}{3} \zeta_5 \Nf^3 \alpha^2
\right. \nonumber \\
&& \left. ~~~
+ \frac{8}{3} \Nf^3 \alpha^3
+ \frac{9}{64} \zeta_3 \Nf \alpha^6
+ \frac{63}{16} \zeta_4 \Nf^2 \alpha^2
+ \frac{315}{64} \zeta_5 \Nf \alpha^6
+ \frac{477}{64} \Nf \alpha^6
\right. \nonumber \\
&& \left. ~~~
+ \frac{621}{64} \zeta_3^2 \Nf \alpha^5
+ \frac{1031}{2} \zeta_3 \Nf^2 \alpha^3
+ \frac{1053}{64} \zeta_4 \Nf \alpha^3
+ \frac{1575}{256} \zeta_6 \Nf \alpha^4
\right. \nonumber \\
&& \left. ~~~
+ \frac{2240}{27} \zeta_3^2 \Nf^3
+ \frac{2961}{4} \zeta_4 \Nf \alpha
+ \frac{3159}{64} \zeta_3^2 \Nf \alpha^3
+ \frac{3549}{1024} \zeta_7 \Nf \alpha^5
\right. \nonumber \\
&& \left. ~~~
+ \frac{3610}{9} \zeta_3 \Nf^3 \alpha
+ \frac{7065}{128} \zeta_3^2 \Nf \alpha^4
+ \frac{9387}{2} \zeta_3^2 \Nf \alpha
+ \frac{14175}{1024} \zeta_6 \alpha^5
\right. \nonumber \\
&& \left. ~~~
+ \frac{19695}{16} \Nf \alpha^4
+ \frac{19791}{128} \Nf \alpha^5
+ \frac{27945}{512} \zeta_3^2 \alpha^6
+ \frac{36525}{256} \zeta_5 \Nf \alpha^4
\right. \nonumber \\
&& \left. ~~~
+ \frac{44145}{512} \zeta_4 \alpha^4
+ \frac{51237}{128} \zeta_3^2 \Nf \alpha^2
+ \frac{66825}{256} \zeta_6 \alpha^3
+ \frac{102619}{36} \zeta_5 \Nf^2 \alpha
\right. \nonumber \\
&& \left. ~~~
+ \frac{114615}{512} \zeta_4 \alpha^2
+ \frac{129545}{162} \zeta_3 \Nf^3
+ \frac{159705}{8192} \zeta_7 \alpha^6
+ \frac{316347}{256} \zeta_5 \Nf \alpha^3
\right. \nonumber \\
&& \left. ~~~
+ \frac{609525}{2048} \zeta_3^2 \alpha^5
+ \frac{2420705}{324} \Nf^3
+ \frac{2646567}{2048} \zeta_3 \alpha^5
+ \frac{2928303}{1024} \zeta_7 \Nf \alpha^2
\right. \nonumber \\
&& \left. ~~~
+ \frac{5923125}{1024} \zeta_6 \alpha
+ \frac{10840253}{1728} \Nf^2 \alpha^2
+ \frac{12012619}{384} \zeta_3 \Nf \alpha^2
\right. \nonumber \\
&& \left. ~~~
+ \frac{13011521}{768} \zeta_5 \Nf \alpha^2
+ \frac{13148441}{2592} \zeta_3 \Nf^2
+ \frac{18387049}{96} \zeta_3 \Nf \alpha
\right. \nonumber \\
&& \left. ~~~
+ \frac{26611431}{512} \zeta_7 \Nf \alpha
+ \frac{27183969}{2048} \zeta_3 \alpha^4
+ \frac{28589639}{3456} \Nf^2 \alpha
\right. \nonumber \\
&& \left. ~~~
+ \frac{36371421}{1024} \zeta_3 \alpha^3
+ \frac{47616093}{8192} \zeta_7 \alpha^3
+ \frac{177532569}{2048} \zeta_5 \alpha^2
\right. \nonumber \\
&& \left. ~~~
+ \frac{246191965}{1296} \zeta_5 \Nf^2
+ \frac{671260095}{2048} \zeta_3^2
+ \frac{1011473605}{3072} \alpha^2
\right. \nonumber \\
&& \left. ~~~
- 48 \zeta_3^2 \Nf^2 \alpha
- 41 \Nf^2 \alpha^4
- 5 \zeta_3 \Nf^3 \alpha^2
+ 3 \zeta_4 \Nf^3 \alpha
+ 15 \zeta_3 \Nf^2 \alpha^4
\right] a^5
\nonumber \\
&& +~ O(a^6)
\end{eqnarray}
\begin{eqnarray}
\left. \gamma_c^{\mMOMs}(a,\alpha) \right|^{SU(3)} &=&
\left[
\frac{3}{4} \alpha
- \frac{9}{4}
\right] a
+ \left[
\frac{3}{4} \Nf
+ \frac{27}{16} \alpha^2
- \frac{153}{8}
- \frac{27}{16} \alpha
\right] a^2
\nonumber \\
&&
+ \left[
\frac{1401}{16} \Nf
- \frac{11691}{16}
- \frac{405}{64} \alpha^3
- \frac{81}{16} \zeta_3 \alpha^2
- \frac{57}{16} \Nf \alpha
- \frac{9}{8} \Nf \alpha^2
- \frac{5}{2} \Nf^2
\right. \nonumber \\
&& \left. ~~~
+ \frac{9}{4} \zeta_3 \Nf
+ \frac{9}{4} \zeta_3 \Nf \alpha
+ \frac{189}{32} \zeta_3 \alpha
+ \frac{243}{8} \alpha^2
+ \frac{243}{32} \zeta_3 \alpha^3
+ \frac{981}{64} \alpha
\right. \nonumber \\
&& \left. ~~~
+ \frac{1269}{16} \zeta_3
\right] a^3
\nonumber \\
&&
+ \left[
\frac{5}{2} \Nf^3
- \frac{16985133}{512}
- \frac{182601}{256} \zeta_3 \alpha
- \frac{158895}{256} \zeta_5 \alpha^2
- \frac{84699}{512} \alpha^3
\right. \nonumber \\
&& \left. ~~~
- \frac{38205}{256} \zeta_5 \alpha^3
- \frac{24183}{128} \Nf \alpha^2
- \frac{18117}{128} \alpha^4
- \frac{16059}{32} \zeta_3 \Nf
\right. \nonumber \\
&& \left. ~~~
- \frac{15693}{64} \Nf \alpha
- \frac{13115}{48} \Nf^2
- \frac{5895}{16} \zeta_5 \Nf
- \frac{585}{8} \zeta_5 \Nf \alpha
- \frac{459}{32} \zeta_3 \Nf \alpha^2
\right. \nonumber \\
&& \left. ~~~
- \frac{27}{8} \zeta_3 \Nf \alpha^3
+ \frac{9}{8} \Nf \alpha^3
+ \frac{27}{64} \Nf \alpha^4
+ \frac{45}{8} \Nf^2 \alpha^2
+ \frac{45}{8} \zeta_3 \Nf \alpha
+ \frac{49}{8} \Nf^2 \alpha
\right. \nonumber \\
&& \left. ~~~
+ \frac{315}{16} \zeta_5 \Nf \alpha^2
+ \frac{729}{128} \zeta_3 \alpha^2
+ \frac{2565}{64} \zeta_5 \alpha^4
+ \frac{31131}{512} \zeta_3 \alpha^4
+ \frac{42417}{256} \zeta_3 \alpha^3
\right. \nonumber \\
&& \left. ~~~
+ \frac{94365}{256} \zeta_5 \alpha
+ \frac{658287}{512} \alpha^2
+ \frac{739053}{128} \Nf
+ \frac{801801}{512} \alpha
+ \frac{1140075}{256} \zeta_5
\right. \nonumber \\
&& \left. ~~~
+ \frac{2264193}{512} \zeta_3
+ 6 \zeta_3 \Nf^2 \alpha
+ 26 \zeta_3 \Nf^2
\right] a^4
\nonumber \\
&&
+ \left[
\frac{9758887695}{8192} \zeta_5
- \frac{15114361941}{32768} \zeta_7
- \frac{9464865363}{4096}
- \frac{458876853}{4096} \zeta_3 \alpha
\right. \nonumber \\
&& \left. ~~~
- \frac{226707579}{4096} \zeta_5 \alpha^2
- \frac{149725879}{9216} \Nf \alpha^2
- \frac{107008841}{3072} \Nf \alpha
\right. \nonumber \\
&& \left. ~~~
- \frac{40486293}{4096} \alpha^3
- \frac{36922473}{16384} \zeta_7 \alpha^3
- \frac{34433025}{256} \zeta_5 \Nf
\right. \nonumber \\
&& \left. ~~~
- \frac{24205473}{4096} \alpha^4
- \frac{23064993}{32768} \zeta_7 \alpha^4
- \frac{14639049}{4096} \zeta_5 \alpha^3
\right. \nonumber \\
&& \left. ~~~
- \frac{12230645}{288} \Nf^2
- \frac{6663195}{2048} \zeta_7 \Nf \alpha
- \frac{1478493}{1024} \alpha^5
- \frac{581013}{512} \zeta_4 \alpha
\right. \nonumber \\
&& \left. ~~~
- \frac{492075}{1024} \zeta_6 \alpha^2
- \frac{397467}{4096} \zeta_3^2 \alpha^4
- \frac{390789}{512} \zeta_7 \Nf \alpha^2
- \frac{144171}{256} \zeta_3^2 \Nf \alpha
\right. \nonumber \\
&& \left. ~~~
- \frac{109683}{256} \zeta_3 \Nf \alpha^3
- \frac{86231}{2} \zeta_3 \Nf
- \frac{66825}{1024} \zeta_6 \alpha^4
- \frac{15795}{1024} \zeta_4 \alpha^3
\right. \nonumber \\
&& \left. ~~~
- \frac{6111}{128} \zeta_3 \Nf \alpha^2
- \frac{4725}{128} \zeta_5 \Nf \alpha^4
- \frac{4131}{1024} \zeta_4 \alpha^5
- \frac{3969}{2} \zeta_7 \Nf^2
\right. \nonumber \\
&& \left. ~~~
- \frac{3645}{8} \zeta_5 \Nf^2 \alpha
- \frac{3013}{64} \zeta_3 \Nf^2 \alpha
- \frac{435}{16} \zeta_5 \Nf^2 \alpha^2
- \frac{405}{128} \zeta_4 \Nf \alpha^3
\right. \nonumber \\
&& \left. ~~~
- \frac{297}{8} \zeta_4 \Nf \alpha^2
- \frac{223}{8} \Nf^3 \alpha
- \frac{219}{2} \zeta_3^2 \Nf^2
- \frac{153}{4} \zeta_3 \Nf^2 \alpha^2
- \frac{135}{32} \Nf^2 \alpha^4
\right. \nonumber \\
&& \left. ~~~
- \frac{81}{8} \zeta_4 \Nf^2 \alpha
- \frac{45}{4} \zeta_3^2 \Nf^2 \alpha^2
- \frac{33}{2} \zeta_3^2 \Nf^2 \alpha
- \frac{27}{2} \zeta_3 \Nf^2 \alpha^3
- \frac{15}{2} \Nf^3 \alpha^2
\right. \nonumber \\
&& \left. ~~~
+ \frac{27}{4} \zeta_3 \Nf \alpha^4
+ \frac{27}{8} \zeta_4 \Nf^2 \alpha^2
+ \frac{81}{64} \zeta_3 \Nf \alpha^5
+ \frac{117}{16} \Nf^2 \alpha^3
+ \frac{297}{256} \Nf \alpha^5
\right. \nonumber \\
&& \left. ~~~
+ \frac{11259}{256} \zeta_3^2 \Nf \alpha^3
+ \frac{14175}{2048} \zeta_6 \alpha^5
+ \frac{15353}{12} \Nf^3
+ \frac{17307}{128} \Nf \alpha^4
\right. \nonumber \\
&& \left. ~~~
+ \frac{17901}{128} \zeta_4 \Nf \alpha
+ \frac{18225}{256} \zeta_6 \alpha^3
+ \frac{29403}{128} \zeta_3^2 \Nf \alpha^2
+ \frac{38637}{1024} \zeta_4 \alpha^4
\right. \nonumber \\
&& \left. ~~~
+ \frac{46935}{512} \zeta_5 \Nf \alpha^3
+ \frac{198531}{1024} \zeta_4 \alpha^2
+ \frac{220563}{4096} \zeta_3^2 \alpha^5
+ \frac{304689}{1024} \Nf \alpha^3
\right. \nonumber \\
&& \left. ~~~
+ \frac{318969}{2048} \zeta_7 \Nf \alpha^3
+ \frac{452709}{2048} \zeta_3^2 \alpha^2
+ \frac{477395}{192} \zeta_3 \Nf^2
+ \frac{550645}{96} \zeta_5 \Nf^2
\right. \nonumber \\
&& \left. ~~~
+ \frac{683829}{256} \zeta_5 \Nf \alpha
+ \frac{687307}{864} \Nf^2 \alpha^2
+ \frac{814149}{128} \zeta_3^2 \Nf
+ \frac{832923}{4096} \zeta_3 \alpha^5
\right. \nonumber \\
&& \left. ~~~
+ \frac{1582659}{2048} \zeta_3^2 \alpha^3
+ \frac{1814049}{512} \zeta_5 \Nf \alpha^2
+ \frac{2473413}{256} \zeta_3 \Nf \alpha
\right. \nonumber \\
&& \left. ~~~
+ \frac{4100625}{2048} \zeta_6 \alpha
+ \frac{4414969}{2304} \Nf^2 \alpha
+ \frac{4890807}{8192} \zeta_5 \alpha^4
+ \frac{5021595}{8192} \zeta_5 \alpha^5
\right. \nonumber \\
&& \left. ~~~
+ \frac{5543559}{32768} \zeta_7 \alpha^5
+ \frac{8992593}{2048} \zeta_3 \alpha^2
+ \frac{13481559}{4096} \zeta_3^2 \alpha
+ \frac{13990239}{4096} \zeta_3 \alpha^4
\right. \nonumber \\
&& \left. ~~~
+ \frac{17310645}{2048} \zeta_3 \alpha^3
+ \frac{34354859}{1024} \zeta_7 \Nf
+ \frac{59929767}{8192} \zeta_5 \alpha
\right. \nonumber \\
&& \left. ~~~
+ \frac{300532491}{16384} \zeta_7 \alpha^2
+ \frac{523935217}{6144} \alpha^2
+ \frac{694222603}{4096} \alpha
+ \frac{1106092719}{4096} \zeta_3
\right. \nonumber \\
&& \left. ~~~
+ \frac{1670942731}{3072} \Nf
+ \frac{1894706667}{32768} \zeta_7 \alpha
- \frac{424982943}{4096} \zeta_3^2
- 65 \zeta_5 \Nf^3
\right. \nonumber \\
&& \left. ~~~
- 14 \Nf^4
+ \zeta_3 \Nf^3
+ \zeta_3 \Nf^3 \alpha
+ 15 \zeta_5 \Nf^3 \alpha
\right] a^5
~+~ O(a^6)
\end{eqnarray}
and
\begin{eqnarray}
\left. \gamma_\psi^{\mMOMs}(a,\alpha) \right|^{SU(3)} &=&
\frac{4}{3} \alpha a
+ \left[
3 \alpha^2
+ 6 \alpha
- \frac{4}{3} \Nf
+ \frac{67}{3}
\right] a^2
\nonumber \\
&&
+ \left[
\frac{29675}{36}
- \frac{706}{9} \Nf
- \frac{607}{2} \zeta_3
- \frac{309}{2} \zeta_3 \alpha
- \frac{45}{2} \zeta_3 \alpha^2
- \frac{8}{3} \Nf \alpha
+ \frac{8}{9} \Nf^2
\right. \nonumber \\
&& \left. ~~~
+ \frac{9}{2} \zeta_3 \alpha^3
+ \frac{49}{12} \alpha^3
+ \frac{107}{4} \alpha^2
+ \frac{427}{4} \alpha
+ 2 \Nf \alpha^2
+ 8 \zeta_3 \Nf \alpha
+ 16 \zeta_3 \Nf
\right] a^3
\nonumber \\
&&
+ \left[
\frac{31003343}{648}
- \frac{21683117}{648} \zeta_3
- \frac{2393555}{324} \Nf
- \frac{51365}{48} \alpha^2
- \frac{17972}{3} \zeta_3 \alpha
\right. \nonumber \\
&& \left. ~~~
- \frac{13087}{54} \Nf \alpha
- \frac{2517}{8} \zeta_3 \alpha^2
- \frac{1841}{72} \alpha^4
- \frac{1640}{9} \zeta_5 \Nf \alpha
- \frac{272}{9} \zeta_3 \Nf^2
\right. \nonumber \\
&& \left. ~~~
- \frac{135}{4} \zeta_5 \alpha^3
- \frac{68}{3} \zeta_3 \Nf \alpha^2
- \frac{40}{9} \Nf^3
- \frac{3}{4} \Nf \alpha^4
+ \frac{23}{2} \Nf \alpha^3
+ \frac{188}{27} \Nf^2 \alpha
\right. \nonumber \\
&& \left. ~~~
+ \frac{235}{16} \zeta_5 \alpha^4
+ \frac{341}{48} \alpha^3
+ \frac{539}{9} \zeta_3 \alpha^4
+ \frac{635}{4} \Nf \alpha^2
+ \frac{2381}{12} \zeta_3 \alpha^3
\right. \nonumber \\
&& \left. ~~~
+ \frac{2861}{9} \Nf^2
+ \frac{3812}{9} \zeta_3 \Nf \alpha
+ \frac{4465}{8} \zeta_5 \alpha^2
+ \frac{34175}{12} \zeta_5 \alpha
+ \frac{74440}{27} \zeta_3 \Nf
\right. \nonumber \\
&& \left. ~~~
+ \frac{194005}{72} \alpha
+ \frac{15846715}{1296} \zeta_5
- 830 \zeta_5 \Nf
- 12 \zeta_3 \Nf \alpha^3
- 10 \zeta_5 \Nf \alpha^2
\right. \nonumber \\
&& \left. ~~~
- 2 \Nf^2 \alpha^2
\right] a^4
\nonumber \\
&&
+ \left[
\frac{2313514793}{10368} \zeta_3^2
- \frac{94958116621}{31104} \zeta_3
- \frac{26588447977}{27648} \zeta_7
\right. \nonumber \\
&& \left. ~~~
+ \frac{14723323093}{5184}
+ \frac{18607183745}{7776} \zeta_5
- \frac{667846415}{1944} \zeta_5 \Nf
\right. \nonumber \\
&& \left. ~~~
- \frac{512366237}{3456} \zeta_3 \alpha
- \frac{409534595}{3456} \alpha^2
- \frac{251804567}{432} \Nf
\right. \nonumber \\
&& \left. ~~~
- \frac{100358063}{3456} \zeta_5 \alpha^2
- \frac{82530427}{768} \zeta_7 \alpha
- \frac{35557175}{432} \zeta_5 \alpha
\right. \nonumber \\
&& \left. ~~~
- \frac{10141489}{1296} \zeta_3 \Nf \alpha^2
- \frac{4726621}{243} \zeta_3 \Nf^2
- \frac{1141313}{144} \alpha^3
- \frac{841917}{128} \zeta_5 \alpha^3
\right. \nonumber \\
&& \left. ~~~
- \frac{596849}{54} \zeta_3^2 \Nf
- \frac{489041}{36} \Nf \alpha
- \frac{455879}{1728} \alpha^5
- \frac{223663}{243} \Nf^2 \alpha^2
\right. \nonumber \\
&& \left. ~~~
- \frac{151875}{128} \zeta_6 \alpha^2
- \frac{137523}{128} \zeta_5 \alpha^4
- \frac{98477}{72} \zeta_3 \Nf \alpha^3
- \frac{80606}{81} \Nf^3
\right. \nonumber \\
&& \left. ~~~
- \frac{72515}{12} \zeta_3^2 \Nf \alpha
- \frac{44073}{128} \zeta_3^2 \alpha^4
- \frac{37779}{1024} \zeta_7 \alpha^4
- \frac{21651}{16} \zeta_3^2 \alpha^3
\right. \nonumber \\
&& \left. ~~~
- \frac{14175}{128} \zeta_6 \alpha^3
- \frac{11996}{27} \zeta_3 \Nf^2 \alpha
- \frac{10125}{128} \zeta_6 \alpha^4
- \frac{6811}{2} \zeta_7 \Nf^2
\right. \nonumber \\
&& \left. ~~~
- \frac{3520}{27} \zeta_5 \Nf^3
- \frac{1617}{32} \zeta_7 \Nf \alpha^3
- \frac{663}{8} \zeta_4 \Nf \alpha^2
- \frac{459}{64} \zeta_4 \alpha^5
- \frac{435}{8} \Nf \alpha^4
\right. \nonumber \\
&& \left. ~~~
- \frac{190}{3} \zeta_5 \Nf^2 \alpha^2
- \frac{139}{6} \Nf^2 \alpha^3
- \frac{129}{4} \zeta_3^2 \Nf \alpha^2
- \frac{128}{9} \zeta_3^2 \Nf^2
- \frac{57}{8} \Nf \alpha^5
\right. \nonumber \\
&& \left. ~~~
- \frac{45}{8} \zeta_4 \Nf \alpha^3
+ \frac{3}{2} \Nf^2 \alpha^4
+ \frac{9}{2} \zeta_3 \Nf \alpha^5
+ \frac{40}{3} \Nf^3 \alpha^2
+ \frac{128}{27} \zeta_3 \Nf^3 \alpha
\right. \nonumber \\
&& \left. ~~~
+ \frac{160}{3} \zeta_5 \Nf^3 \alpha
+ \frac{160}{27} \Nf^4
+ \frac{225}{8} \zeta_5 \Nf \alpha^4
+ \frac{413}{24} \zeta_7 \Nf \alpha^2
+ \frac{729}{16} \zeta_4 \alpha^4
\right. \nonumber \\
&& \left. ~~~
+ \frac{1575}{128} \zeta_6 \alpha^5
+ \frac{2403}{128} \zeta_3^2 \alpha^5
+ \frac{3455}{8} \zeta_5 \Nf \alpha^3
+ \frac{6993}{64} \zeta_4 \alpha^3
\right. \nonumber \\
&& \left. ~~~
+ \frac{8083}{81} \zeta_3 \Nf^2 \alpha^2
+ \frac{14931}{256} \zeta_7 \alpha^5
+ \frac{19490}{27} \Nf^2 \alpha
+ \frac{21519}{32} \zeta_4 \alpha^2
\right. \nonumber \\
&& \left. ~~~
+ \frac{24064}{81} \zeta_3 \Nf^3
+ \frac{28951}{4} \zeta_3^2 \alpha^2
+ \frac{33825}{128} \zeta_5 \alpha^5
+ \frac{82327}{24} \zeta_5 \Nf \alpha^2
\right. \nonumber \\
&& \left. ~~~
+ \frac{105389}{384} \zeta_3 \alpha^5
+ \frac{123931}{96} \Nf \alpha^3
+ \frac{292055}{1152} \alpha^4
+ \frac{414431}{128} \zeta_3 \alpha^4
\right. \nonumber \\
&& \left. ~~~
+ \frac{442155}{128} \zeta_7 \alpha^3
+ \frac{641081}{96} \zeta_7 \Nf \alpha
+ \frac{817061}{1536} \zeta_7 \alpha^2
+ \frac{1030331}{72} \zeta_3 \alpha^3
\right. \nonumber \\
&& \left. ~~~
+ \frac{1063237}{27} \Nf^2
+ \frac{1295257}{72} \zeta_3 \Nf \alpha
+ \frac{2129515}{72} \zeta_5 \Nf \alpha
+ \frac{3085750}{243} \zeta_5 \Nf^2
\right. \nonumber \\
&& \left. ~~~
+ \frac{4429579}{36} \zeta_7 \Nf
+ \frac{13408341}{128} \zeta_3^2 \alpha
+ \frac{31261961}{432} \zeta_3 \alpha^2
\right. \nonumber \\
&& \left. ~~~
+ \frac{52889729}{2592} \Nf \alpha^2
+ \frac{138867799}{1728} \alpha
+ \frac{1748707267}{3888} \zeta_3 \Nf
- 2475 \zeta_5 \Nf^2 \alpha
\right. \nonumber \\
&& \left. ~~~
- 41 \zeta_3 \Nf \alpha^4
- 8 \Nf^3 \alpha
- 8 \zeta_3^2 \Nf^2 \alpha^2
+ 6 \zeta_4 \Nf^2 \alpha^2
+ 24 \zeta_3^2 \Nf^2 \alpha
\right. \nonumber \\
&& \left. ~~~
+ 128 \zeta_3^2 \Nf \alpha^3
\right] a^5 ~+~ O(a^6)
\end{eqnarray}
for the three fields of the Lagrangian. The $\mMOM$ scheme quark mass anomalous
dimension is
\begin{eqnarray}
\left. \gamma_m^{\mMOMs}(a,\alpha) \right|^{SU(3)} &=&
-~ 4 a
+ \left[
\alpha^2
- \frac{209}{3}
+ \frac{4}{3} \Nf
\right] a^2
\nonumber \\
&&
+ \left[
\frac{5635}{6} \zeta_3
- \frac{95383}{36}
- \frac{176}{9} \zeta_3 \Nf
- \frac{27}{2} \zeta_3 \alpha^2
- \frac{8}{3} \Nf^2
+ \frac{23}{12} \alpha^3
+ \frac{47}{4} \alpha
\right. \nonumber \\
&& \left. ~~~
+ \frac{459}{4} \alpha^2
+ \frac{4742}{27} \Nf
- 2 \Nf \alpha^2
+ 2 \Nf \alpha
+ 99 \zeta_3 \alpha
\right] a^3
\nonumber \\
&&
+ \left[
\frac{8}{3} \Nf^3
- \frac{182707879}{1296}
- \frac{309295}{48} \zeta_5
- \frac{159817}{27} \zeta_3 \Nf
- \frac{47377}{18} \zeta_3 \alpha^2
\right. \nonumber \\
&& \left. ~~~
- \frac{13651}{27} \Nf^2
- \frac{7965}{4} \zeta_5 \alpha
- \frac{6821}{144} \alpha^4
- \frac{4202}{9} \zeta_3 \Nf \alpha
- \frac{3200}{9} \zeta_5 \Nf
\right. \nonumber \\
&& \left. ~~~
- \frac{3007}{12} \zeta_3 \alpha^3
- \frac{2413}{6} \Nf \alpha^2
- \frac{1575}{8} \zeta_5 \alpha^2
- \frac{176}{27} \Nf^2 \alpha
- \frac{63}{8} \zeta_3 \alpha^4
\right. \nonumber \\
&& \left. ~~~
- \frac{45}{4} \zeta_5 \alpha^3
- \frac{21}{2} \Nf \alpha^3
+ \frac{3}{4} \Nf \alpha^4
+ \frac{315}{16} \zeta_5 \alpha^4
+ \frac{608}{27} \zeta_3 \Nf^2 \alpha
\right. \nonumber \\
&& \left. ~~~
+ \frac{1019}{108} \Nf \alpha
+ \frac{1265}{27} \zeta_3 \Nf \alpha^2
+ \frac{1552}{9} \zeta_3 \Nf^2
+ \frac{10571}{24} \alpha^3
+ \frac{12541}{4} \zeta_3 \alpha
\right. \nonumber \\
&& \left. ~~~
+ \frac{333899}{48} \alpha^2
+ \frac{480463}{144} \alpha
+ \frac{5246557}{324} \Nf
+ \frac{15752321}{216} \zeta_3
+ 6 \Nf^2 \alpha^2
\right] a^4
\nonumber \\
&&
+ \left[
\frac{3576071485}{27648} \zeta_7
- \frac{75504232175}{7776}
+ \frac{9610932889}{5832} \Nf
+ \frac{17917034005}{31104} \zeta_5
\right. \nonumber \\
&& \left. ~~~
+ \frac{187324052147}{31104} \zeta_3
- \frac{448693433}{1728} \zeta_3 \alpha^2
- \frac{310328447}{432} \zeta_3^2
\right. \nonumber \\
&& \left. ~~~
- \frac{257106335}{324} \zeta_3 \Nf
- \frac{180251015}{1944} \zeta_5 \Nf
- \frac{135482155}{2592} \Nf \alpha^2
\right. \nonumber \\
&& \left. ~~~
- \frac{51199769}{384} \zeta_5 \alpha
- \frac{22459484}{243} \Nf^2
- \frac{13215491}{576} \zeta_3 \alpha^3
- \frac{9488839}{288} \Nf \alpha
\right. \nonumber \\
&& \left. ~~~
- \frac{6906043}{1152} \zeta_3 \alpha
- \frac{4778536}{81} \zeta_7 \Nf
- \frac{1261649}{16} \zeta_3^2 \alpha
- \frac{540175}{144} \alpha^4
\right. \nonumber \\
&& \left. ~~~
- \frac{421721}{864} \alpha^5
- \frac{363001}{36} \zeta_3 \Nf \alpha
- \frac{126619}{48} \zeta_5 \alpha^2
- \frac{75565}{48} \Nf \alpha^3
\right. \nonumber \\
&& \left. ~~~
- \frac{64557}{32} \zeta_4 \alpha
- \frac{60928}{81} \zeta_3^2 \Nf^2
- \frac{47287}{81} \zeta_3 \Nf^2 \alpha^2
- \frac{28096}{81} \zeta_3 \Nf^3
\right. \nonumber \\
&& \left. ~~~
- \frac{26173}{8} \zeta_7 \Nf \alpha
- \frac{20979}{64} \zeta_4 \alpha^2
- \frac{17307}{64} \zeta_3^2 \alpha^4
- \frac{10647}{1024} \zeta_7 \alpha^5
\right. \nonumber \\
&& \left. ~~~
- \frac{4725}{128} \zeta_6 \alpha^4
- \frac{2187}{16} \zeta_4 \alpha^3
- \frac{1863}{64} \zeta_3^2 \alpha^5
- \frac{1600}{9} \zeta_5 \Nf^3
- \frac{830}{3} \zeta_5 \Nf^2 \alpha
\right. \nonumber \\
&& \left. ~~~
- \frac{617}{18} \zeta_3 \Nf \alpha^4
- \frac{603}{2} \zeta_3^2 \Nf \alpha^2
- \frac{352}{27} \Nf^4
- \frac{255}{8} \zeta_5 \Nf \alpha^3
- \frac{175}{16} \zeta_7 \Nf \alpha^2
\right. \nonumber \\
&& \left. ~~~
- \frac{152}{3} \zeta_3 \Nf^2 \alpha^3
- \frac{128}{27} \zeta_3 \Nf^3 \alpha
- \frac{105}{8} \zeta_5 \Nf \alpha^4
- \frac{10}{3} \zeta_5 \Nf^2 \alpha^2
- \frac{9}{2} \Nf^2 \alpha^4
\right. \nonumber \\
&& \left. ~~~
+ \frac{27}{4} \Nf \alpha^5
+ \frac{124}{9} \Nf^3 \alpha
+ \frac{135}{8} \zeta_4 \Nf \alpha^2
+ \frac{223}{6} \Nf^2 \alpha^3
+ \frac{1037}{4} \Nf \alpha^4
\right. \nonumber \\
&& \left. ~~~
+ \frac{1372}{3} \zeta_7 \Nf^2
+ \frac{1377}{64} \zeta_4 \alpha^4
+ \frac{1989}{8} \zeta_4 \Nf \alpha
+ \frac{7639}{9} \zeta_3 \Nf^2 \alpha
\right. \nonumber \\
&& \left. ~~~
+ \frac{10435}{6} \zeta_3^2 \Nf \alpha
+ \frac{13915}{384} \zeta_3 \alpha^5
+ \frac{15425}{8} \zeta_5 \Nf \alpha^2
+ \frac{21784}{81} \Nf^2 \alpha
\right. \nonumber \\
&& \left. ~~~
+ \frac{27735}{128} \zeta_5 \alpha^5
+ \frac{30375}{128} \zeta_6 \alpha^3
+ \frac{42525}{128} \zeta_6 \alpha^2
+ \frac{46027}{36} \zeta_3 \Nf \alpha^3
\right. \nonumber \\
&& \left. ~~~
+ \frac{64065}{128} \zeta_5 \alpha^4
+ \frac{92475}{64} \zeta_3^2 \alpha^3
+ \frac{95931}{32} \zeta_5 \alpha^3
+ \frac{123604}{9} \zeta_5 \Nf \alpha
\right. \nonumber \\
&& \left. ~~~
+ \frac{253453}{162} \Nf^2 \alpha^2
+ \frac{316953}{512} \zeta_7 \alpha^2
+ \frac{336231}{1024} \zeta_7 \alpha^4
+ \frac{439047}{512} \zeta_7 \alpha^3
\right. \nonumber \\
&& \left. ~~~
+ \frac{455039}{384} \zeta_3 \alpha^4
+ \frac{455625}{128} \zeta_6 \alpha
+ \frac{464038}{243} \Nf^3
+ \frac{466231}{64} \zeta_3^2 \alpha^2
\right. \nonumber \\
&& \left. ~~~
+ \frac{948548}{27} \zeta_3 \Nf^2
+ \frac{1850845}{243} \zeta_5 \Nf^2
+ \frac{6570181}{162} \zeta_3^2 \Nf
+ \frac{6589163}{324} \zeta_3 \Nf \alpha^2
\right. \nonumber \\
&& \left. ~~~
+ \frac{31369981}{1152} \alpha^3
+ \frac{61494153}{1024} \zeta_7 \alpha
+ \frac{71828867}{144} \alpha^2
+ \frac{1414667209}{3456} \alpha
\right. \nonumber \\
&& \left. ~~~
- 18 \zeta_4 \Nf^2 \alpha
- 9 \zeta_3^2 \Nf \alpha^3
- 8 \Nf^3 \alpha^2
\right] a^5 ~+~ O(a^6) ~.
\end{eqnarray}
The respective conversion functions that these are derived from are
\begin{eqnarray}
\left. C_A(a,\alpha) \right|^{SU(3)} &=& 1
+ \left[
\frac{3}{2} \alpha
+ \frac{3}{4} \alpha^2
+ \frac{97}{12}
- \frac{10}{9} \Nf
\right] a
\nonumber \\
&&
+ \left[
\frac{83105}{288}
- \frac{11299}{216} \Nf
- \frac{5}{3} \Nf \alpha
- \frac{4}{3} \zeta_3 \Nf
+ \frac{45}{16} \alpha^3
+ \frac{100}{81} \Nf^2
+ \frac{135}{16} \alpha^2
\right. \nonumber \\
&& \left. ~~~
+ \frac{313}{32} \alpha
- 27 \zeta_3
+ 18 \zeta_3 \alpha
\right] a^2
\nonumber \\
&&
+ \left[
\frac{164395}{972} \Nf^2
- \frac{8228977}{2592} \Nf
- \frac{63225}{64} \zeta_5
- \frac{17433}{8} \zeta_3
- \frac{8821}{144} \Nf \alpha
\right. \nonumber \\
&& \left. ~~~
- \frac{3465}{32} \zeta_5 \alpha^2
- \frac{3105}{8} \zeta_5 \alpha
- \frac{1000}{729} \Nf^3
- \frac{477}{32} \zeta_3 \alpha^3
- \frac{469}{32} \Nf \alpha^2
- \frac{243}{32} \zeta_4
\right. \nonumber \\
&& \left. ~~~
- \frac{117}{32} \zeta_3 \alpha^4
- \frac{81}{8} \zeta_4 \alpha
- \frac{81}{32} \zeta_4 \alpha^2
+ \frac{3}{4} \zeta_3 \Nf \alpha^2
+ \frac{14}{27} \Nf^2 \alpha
+ \frac{16}{3} \zeta_3 \Nf^2 \alpha
\right. \nonumber \\
&& \left. ~~~
+ \frac{16}{9} \zeta_3 \Nf^2
+ \frac{33}{2} \zeta_4 \Nf
+ \frac{45}{8} \zeta_5 \alpha^3
+ \frac{315}{64} \zeta_5 \alpha^4
+ \frac{1197}{64} \alpha^4
\right. \nonumber \\
&& \left. ~~~
+ \frac{1881}{32} \zeta_3 \alpha^2
+ \frac{2320}{9} \zeta_5 \Nf
+ \frac{11961}{128} \alpha^3
+ \frac{25915}{108} \zeta_3 \Nf
+ \frac{40539}{128} \alpha^2
\right. \nonumber \\
&& \left. ~~~
+ \frac{42933}{32} \zeta_3 \alpha
+ \frac{161965}{384} \alpha
+ \frac{44961125}{3456}
- 161 \zeta_3 \Nf \alpha
\right] a^3
\nonumber \\
&&
+ \left[
\frac{62302764631}{82944}
- \frac{1758762815}{7776} \Nf
- \frac{324121925}{2048} \zeta_3
- \frac{262747689}{1024} \zeta_5
\right. \nonumber \\
&& \left. ~~~
- \frac{62931331}{13824} \Nf \alpha
- \frac{41645709}{1024} \zeta_5 \alpha
- \frac{25666081}{864} \zeta_7 \Nf
- \frac{15059695}{34992} \Nf^3
\right. \nonumber \\
&& \left. ~~~
- \frac{8797395}{1024} \zeta_5 \alpha^2
- \frac{4480985}{288} \zeta_3 \Nf \alpha
- \frac{3319667}{2304} \Nf \alpha^2
- \frac{3161781}{2048} \zeta_4
\right. \nonumber \\
&& \left. ~~~
- \frac{826983}{1024} \zeta_4 \alpha
- \frac{515843}{162} \zeta_5 \Nf^2
- \frac{487965}{1024} \zeta_3 \alpha^3
- \frac{468495}{2048} \zeta_3 \alpha^4
\right. \nonumber \\
&& \left. ~~~
- \frac{362555}{648} \zeta_3 \Nf^2
- \frac{174483}{32} \zeta_3^2
- \frac{170451}{512} \zeta_3^2 \alpha
- \frac{132615}{512} \zeta_7 \alpha^3
\right. \nonumber \\
&& \left. ~~~
- \frac{61911}{512} \zeta_4 \alpha^2
- \frac{26901}{32} \zeta_7 \Nf \alpha
- \frac{24975}{1024} \zeta_6 \alpha^3
- \frac{23925}{64} \zeta_3 \Nf \alpha^2
\right. \nonumber \\
&& \left. ~~~
- \frac{22275}{1024} \zeta_6 \alpha^2
- \frac{19737}{512} \zeta_3^2 \alpha^4
- \frac{16775}{16} \zeta_6 \Nf
- \frac{14985}{512} \zeta_3 \alpha^5
- \frac{10647}{4096} \zeta_7 \alpha^5
\right. \nonumber \\
&& \left. ~~~
- \frac{9621}{64} \Nf \alpha^3
- \frac{4725}{1024} \zeta_6 \alpha^4
- \frac{1943}{216} \zeta_4 \Nf^2
- \frac{1863}{256} \zeta_3^2 \alpha^5
- \frac{1107}{8} \zeta_3^2 \Nf
\right. \nonumber \\
&& \left. ~~~
- \frac{995}{32} \zeta_5 \Nf \alpha^3
- \frac{513}{4} \zeta_3^2 \Nf \alpha
- \frac{229}{27} \zeta_3 \Nf^3
- \frac{160}{9} \zeta_3 \Nf^3 \alpha
- \frac{115}{2} \zeta_5 \Nf^2 \alpha
\right. \nonumber \\
&& \left. ~~~
- \frac{81}{1024} \zeta_4 \alpha^3
- \frac{63}{8} \zeta_4 \Nf \alpha^2
- \frac{45}{4} \zeta_3^2 \Nf \alpha^2
- \frac{9}{4} \zeta_4 \Nf^2 \alpha
- \frac{9}{4} \zeta_3^2 \Nf \alpha^3
\right. \nonumber \\
&& \left. ~~~
+ \frac{5}{2} \zeta_5 \Nf^2 \alpha^2
+ \frac{83}{6} \zeta_3 \Nf^2 \alpha^2
+ \frac{279}{4} \zeta_4 \Nf \alpha
+ \frac{400}{9} \zeta_3^2 \Nf^2
+ \frac{580}{243} \Nf^3 \alpha
\right. \nonumber \\
&& \left. ~~~
+ \frac{880}{27} \zeta_5 \Nf^3
+ \frac{2879}{3} \zeta_3 \Nf^2 \alpha
+ \frac{5265}{2048} \zeta_4 \alpha^4
+ \frac{6647}{32} \zeta_5 \Nf \alpha^2
+ \frac{6781}{432} \Nf^2 \alpha^2
\right. \nonumber \\
&& \left. ~~~
+ \frac{10000}{6561} \Nf^4
+ \frac{28575}{256} \zeta_5 \alpha^4
+ \frac{31635}{512} \zeta_5 \alpha^5
+ \frac{37719}{256} \alpha^5
+ \frac{56025}{512} \zeta_3^2 \alpha^3
\right. \nonumber \\
&& \left. ~~~
+ \frac{137781}{1024} \zeta_7 \alpha^4
+ \frac{156711}{1024} \zeta_5 \alpha^3
+ \frac{204525}{32} \zeta_6
+ \frac{373855}{96} \zeta_5 \Nf \alpha
\right. \nonumber \\
&& \left. ~~~
+ \frac{392337}{512} \zeta_3^2 \alpha^2
+ \frac{469333}{576} \zeta_4 \Nf
+ \frac{651031}{5184} \Nf^2 \alpha
+ \frac{851175}{1024} \zeta_6 \alpha
\right. \nonumber \\
&& \left. ~~~
+ \frac{2124351}{2048} \alpha^4
+ \frac{2297709}{512} \alpha^3
+ \frac{2371053}{512} \zeta_3 \alpha^2
+ \frac{2566431}{2048} \zeta_7 \alpha^2
\right. \nonumber \\
&& \left. ~~~
+ \frac{27270475}{864} \zeta_3 \Nf
+ \frac{49026385}{3072} \alpha^2
+ \frac{60587905}{864} \zeta_5 \Nf
+ \frac{70979391}{4096} \zeta_7 \alpha
\right. \nonumber \\
&& \left. ~~~
+ \frac{78128919}{1024} \zeta_3 \alpha
+ \frac{190787741}{10368} \Nf^2
+ \frac{277127487}{2048} \zeta_7
+ \frac{457175255}{18432} \alpha
\right. \nonumber \\
&& \left. ~~~
- \zeta_4 \Nf^3
+ 25 \zeta_3 \Nf \alpha^3
\right] a^4 ~+~ O(a^5)
\end{eqnarray}
\begin{eqnarray}
\left. C_c(a,\alpha) \right|^{SU(3)} &=& 1 + 3 a
+ \left[
\frac{5829}{64}
- \frac{135}{16} \zeta_3
- \frac{95}{16} \Nf
- \frac{63}{64} \alpha
- \frac{27}{16} \zeta_3 \alpha^2
+ \frac{27}{8} \alpha^2
+ \frac{27}{8} \zeta_3 \alpha
\right] a^2
\nonumber \\
&&
+ \left[
\frac{5161}{648} \Nf^2
- \frac{198001}{432} \Nf
- \frac{40449}{64} \zeta_3
- \frac{3159}{64} \zeta_3 \alpha^2
- \frac{1755}{32} \zeta_5
- \frac{1755}{32} \zeta_5 \alpha
\right. \nonumber \\
&& \left. ~~~
- \frac{621}{64} \zeta_3 \alpha^3
- \frac{195}{64} \alpha
- \frac{135}{32} \zeta_5 \alpha^3
- \frac{39}{4} \zeta_3 \Nf \alpha
- \frac{33}{4} \zeta_4 \Nf
+ \frac{2}{3} \zeta_3 \Nf^2
\right. \nonumber \\
&& \left. ~~~
+ \frac{59}{2} \zeta_3 \Nf
+ \frac{81}{16} \zeta_4 \alpha
+ \frac{81}{64} \zeta_4 \alpha^2
+ \frac{199}{32} \Nf \alpha
+ \frac{243}{64} \zeta_4
+ \frac{945}{32} \zeta_5 \alpha^2
\right. \nonumber \\
&& \left. ~~~
+ \frac{3249}{128} \alpha^3
+ \frac{11421}{128} \alpha^2
+ \frac{14157}{64} \zeta_3 \alpha
+ \frac{1082353}{288}
\right] a^3
\nonumber \\
&&
+ \left[
\frac{15567976783}{73728}
- \frac{151911987}{4096} \zeta_3
- \frac{79190001}{2048} \zeta_5
- \frac{39621021}{1024} \Nf
\right. \nonumber \\
&& \left. ~~~
- \frac{7341417}{2048} \zeta_5 \alpha
- \frac{3869019}{4096} \zeta_7 \alpha
- \frac{3508811}{2304} \alpha
- \frac{1282797}{2048} \zeta_3 \alpha^3
\right. \nonumber \\
&& \left. ~~~
- \frac{1172367}{2048} \zeta_7 \alpha^2
- \frac{455517}{1024} \zeta_3^2 \alpha
- \frac{317277}{4096} \zeta_3 \alpha^4
- \frac{289845}{128} \zeta_3 \alpha^2
\right. \nonumber \\
&& \left. ~~~
- \frac{214893}{8192} \zeta_7 \alpha^4
- \frac{204525}{64} \zeta_6
- \frac{173743}{10368} \Nf^2 \alpha
- \frac{158493}{128} \zeta_3 \Nf \alpha
- \frac{150979}{7776} \Nf^3
\right. \nonumber \\
&& \left. ~~~
- \frac{92113}{512} \Nf \alpha^2
- \frac{83943}{2048} \zeta_5 \alpha^3
- \frac{52025}{128} \zeta_4 \Nf
- \frac{36045}{1024} \zeta_5 \alpha^4
- \frac{27135}{2048} \zeta_4 \alpha^3
\right. \nonumber \\
&& \left. ~~~
- \frac{25947}{1024} \zeta_3^2 \alpha^3
- \frac{11907}{16} \zeta_7 \Nf
- \frac{4725}{2048} \zeta_6 \alpha^4
- \frac{699}{8} \zeta_5 \Nf^2
- \frac{633}{16} \zeta_5 \Nf \alpha^2
\right. \nonumber \\
&& \left. ~~~
- \frac{425}{48} \zeta_3 \Nf^2
- \frac{135}{16} \zeta_3^2 \Nf \alpha^2
- \frac{45}{4} \zeta_3^2 \Nf \alpha
- \frac{9}{8} \zeta_4 \Nf^2 \alpha
- \frac{5}{18} \zeta_3 \Nf^3
+ \frac{1}{2} \zeta_4 \Nf^3
\right. \nonumber \\
&& \left. ~~~
+ \frac{15}{2} \zeta_5 \Nf^2 \alpha
+ \frac{81}{64} \zeta_4 \Nf \alpha^2
+ \frac{117}{64} \zeta_4 \Nf \alpha
+ \frac{153}{8} \zeta_3 \Nf^2 \alpha
+ \frac{165}{16} \zeta_4 \Nf^2
\right. \nonumber \\
&& \left. ~~~
+ \frac{337}{4} \zeta_3^2 \Nf
+ \frac{1161}{1024} \zeta_3^2 \alpha^4
+ \frac{6033}{64} \zeta_5 \Nf \alpha
+ \frac{9639}{4096} \zeta_4 \alpha^4
+ \frac{16281}{1024} \zeta_4 \alpha^2
\right. \nonumber \\
&& \left. ~~~
+ \frac{16775}{32} \zeta_6 \Nf
+ \frac{27459}{256} \zeta_3 \Nf \alpha^2
+ \frac{60075}{2048} \zeta_6 \alpha
+ \frac{85725}{2048} \zeta_6 \alpha^3
+ \frac{107325}{2048} \zeta_6 \alpha^2
\right. \nonumber \\
&& \left. ~~~
+ \frac{130977}{1024} \zeta_3^2 \alpha^2
+ \frac{250085}{64} \zeta_5 \Nf
+ \frac{257823}{2048} \zeta_4 \alpha
+ \frac{956907}{4096} \zeta_7 \alpha^3
+ \frac{1007289}{4096} \alpha^4
\right. \nonumber \\
&& \left. ~~~
+ \frac{1149471}{512} \zeta_3^2
+ \frac{1279017}{1024} \alpha^3
+ \frac{1780139}{3456} \Nf \alpha
+ \frac{2625583}{768} \zeta_3 \Nf
\right. \nonumber \\
&& \left. ~~~
+ \frac{3540987}{2048} \zeta_5 \alpha^2
+ \frac{5819877}{4096} \zeta_4
+ \frac{8093153}{4608} \Nf^2
+ \frac{29075517}{2048} \zeta_3 \alpha
\right. \nonumber \\
&& \left. ~~~
+ \frac{35695263}{8192} \alpha^2
+ \frac{100880073}{8192} \zeta_7
\right] a^4 ~+~ O(a^5)
\end{eqnarray}
and
\begin{eqnarray}
\left. C_\psi(a,\alpha) \right|^{SU(3)} &=& 1 - \frac{4}{3} \alpha a
+ \left[
\frac{7}{3} \Nf
+ 12 \zeta_3
+ 12 \zeta_3 \alpha
- \frac{359}{9}
- \frac{49}{18} \alpha^2
- 26 \alpha
\right] a^2
\nonumber \\
&&
+ \left[
\frac{24722}{81} \Nf
- \frac{439543}{162}
- \frac{322351}{432} \alpha
- \frac{4139}{48} \alpha^2
- \frac{1570}{243} \Nf^2
- \frac{1165}{3} \zeta_5
\right. \nonumber \\
&& \left. ~~~
- \frac{929}{54} \alpha^3
- \frac{440}{9} \zeta_3 \Nf
- \frac{410}{3} \zeta_5 \alpha
- \frac{9}{2} \zeta_4 \alpha
- \frac{9}{4} \zeta_4 \alpha^2
+ \frac{20}{81} \zeta_3 \alpha^3
+ \frac{53}{2} \zeta_3 \alpha^2
\right. \nonumber \\
&& \left. ~~~
+ \frac{79}{4} \zeta_4
+ \frac{1436}{3} \zeta_3 \alpha
+ \frac{1547}{36} \Nf \alpha
+ \frac{8009}{6} \zeta_3
- 16 \zeta_3 \Nf \alpha
- 15 \zeta_5 \alpha^2
\right] a^3
\nonumber \\
&&
+ \left[
\frac{21391}{1458} \Nf^3
- \frac{2065961635}{62208} \alpha
- \frac{356864009}{5184} \zeta_5
- \frac{146722043}{864}
- \frac{74862851}{20736} \alpha^2
\right. \nonumber \\
&& \left. ~~~
- \frac{29889697}{5184} \zeta_3^2
- \frac{1294381}{108} \zeta_3 \Nf
- \frac{1276817}{972} \Nf^2
- \frac{326689}{2592} \alpha^4
\right. \nonumber \\
&& \left. ~~~
- \frac{240893}{288} \zeta_5 \alpha^2
- \frac{171487}{144} \zeta_5 \alpha
- \frac{106555}{128} \alpha^3
- \frac{94613}{1458} \Nf^2 \alpha
- \frac{85687}{192} \zeta_4 \alpha
\right. \nonumber \\
&& \left. ~~~
- \frac{11771}{27} \zeta_5 \Nf \alpha
- \frac{9179}{4} \zeta_3^2 \alpha
- \frac{4851}{64} \zeta_7 \alpha^3
- \frac{1913}{108} \zeta_3 \Nf \alpha^2
- \frac{1437}{16} \zeta_4 \alpha^2
\right. \nonumber \\
&& \left. ~~~
- \frac{757}{16} \zeta_5 \alpha^3
- \frac{685}{64} \zeta_5 \alpha^4
- \frac{501}{64} \zeta_4 \alpha^3
- \frac{440}{9} \zeta_5 \Nf^2
- \frac{275}{128} \zeta_6 \alpha^4
- \frac{20}{3} \zeta_4 \Nf^2
\right. \nonumber \\
&& \left. ~~~
+ \frac{3}{4} \zeta_4 \alpha^4
+ \frac{4}{3} \zeta_3^2 \Nf \alpha
+ \frac{8}{27} \zeta_3 \Nf^3
+ \frac{21}{8} \zeta_4 \Nf \alpha^2
+ \frac{39}{4} \zeta_3^2 \alpha^3
+ \frac{76}{9} \zeta_3 \Nf^2 \alpha
\right. \nonumber \\
&& \left. ~~~
+ \frac{80}{3} \zeta_5 \Nf^2 \alpha
+ \frac{100}{3} \zeta_6 \Nf
+ \frac{149}{12} \zeta_5 \Nf \alpha^2
+ \frac{155}{64} \zeta_3^2 \alpha^4
+ \frac{245}{4} \zeta_4 \Nf \alpha
\right. \nonumber \\
&& \left. ~~~
+ \frac{413}{32} \zeta_7 \alpha^2
+ \frac{2291}{72} \zeta_4 \Nf
+ \frac{3725}{64} \zeta_6 \alpha^2
+ \frac{4651}{32} \zeta_3^2 \alpha^2
+ \frac{5704}{27} \zeta_3 \Nf^2
\right. \nonumber \\
&& \left. ~~~
+ \frac{14801}{72} \zeta_3 \alpha^3
+ \frac{32173}{3888} \zeta_3 \alpha^4
+ \frac{119995}{864} \Nf \alpha^2
+ \frac{387653}{192} \zeta_7 \alpha
+ \frac{565939}{864} \zeta_4
\right. \nonumber \\
&& \left. ~~~
+ \frac{1132909}{864} \zeta_3 \alpha^2
+ \frac{1460149}{72} \zeta_3 \alpha
+ \frac{1673051}{324} \zeta_5 \Nf
+ \frac{3807625}{10368} \zeta_6
\right. \nonumber \\
&& \left. ~~~
+ \frac{6747755}{288} \zeta_7
+ \frac{55476671}{1944} \Nf
+ \frac{59611205}{15552} \Nf \alpha
+ \frac{317781451}{2592} \zeta_3
\right. \nonumber \\
&& \left. ~~~
- 1552 \zeta_3 \Nf \alpha
- 1029 \zeta_7 \Nf
- 24 \zeta_3^2 \Nf
- 6 \zeta_3^2 \Nf \alpha^2
- 2 \zeta_4 \Nf^2 \alpha
+ 575 \zeta_6 \alpha
\right] a^4
\nonumber \\
&& +~ O(a^5) 
\end{eqnarray}
for the field conversion functions while we have  
\begin{eqnarray}
\left. C_m(a,\alpha) \right|^{SU(3)} &=& 1
- \left[
\frac{16}{3}
+ \frac{4}{3} \alpha
\right] a
+ \left[
\frac{83}{9} \Nf
+ \frac{152}{3} \zeta_3
- \frac{3779}{18}
- \frac{62}{9} \alpha
- \frac{11}{9} \alpha^2
\right] a^2
\nonumber \\
&&
+ \left[
\frac{217390}{243} \Nf
- \frac{3115807}{324}
- \frac{36235}{216} \alpha
- \frac{12695}{216} \alpha^2
- \frac{7514}{729} \Nf^2
- \frac{4720}{27} \zeta_3 \Nf
\right. \nonumber \\
&& \left. ~~~
- \frac{2960}{9} \zeta_5
- \frac{1291}{108} \alpha^3
- \frac{761}{9} \zeta_3 \alpha
- \frac{32}{27} \zeta_3 \Nf^2
- \frac{8}{9} \zeta_3 \Nf \alpha
+ \frac{80}{3} \zeta_4 \Nf
\right. \nonumber \\
&& \left. ~~~
+ \frac{211}{18} \zeta_3 \alpha^2
+ \frac{815}{54} \Nf \alpha
+ \frac{195809}{54} \zeta_3
\right] a^3
\nonumber \\
&&
+ \left[
\frac{40}{81} \zeta_3 \Nf^3
- \frac{744609145}{1296}
- \frac{52383125}{17496} \Nf^2
- \frac{26904209}{3888} \alpha
- \frac{4767911}{648} \zeta_3 \alpha
\right. \nonumber \\
&& \left. ~~~
- \frac{3837631}{1728} \zeta_7
- \frac{3053677}{1296} \alpha^2
- \frac{2017309}{81} \zeta_3 \Nf
- \frac{843077}{54} \zeta_3^2
- \frac{711149}{1296} \alpha^3
\right. \nonumber \\
&& \left. ~~~
- \frac{359855}{81} \zeta_5 \Nf
- \frac{115433}{1296} \alpha^4
- \frac{77737}{2187} \Nf^2 \alpha
- \frac{16960}{9} \zeta_4
- \frac{11500}{9} \zeta_6 \Nf
\right. \nonumber \\
&& \left. ~~~
- \frac{8776}{27} \zeta_3^2 \Nf
- \frac{3973}{12} \zeta_3^2 \alpha
- \frac{3155}{8} \zeta_5 \alpha^2
- \frac{2072}{81} \zeta_3 \Nf \alpha^2
- \frac{343}{2} \zeta_7 \Nf
\right. \nonumber \\
&& \left. ~~~
- \frac{122}{9} \zeta_4 \Nf \alpha
- \frac{100}{3} \zeta_4 \Nf^2
- \frac{91}{2} \zeta_7 \alpha
- \frac{81}{8} \zeta_4 \alpha
- \frac{80}{27} \zeta_3 \Nf^2 \alpha
- \frac{55}{16} \zeta_5 \alpha^4
\right. \nonumber \\
&& \left. ~~~
- \frac{27}{2} \zeta_4 \alpha^2
- \frac{27}{4} \zeta_3^2 \alpha^2
- \frac{27}{8} \zeta_4 \alpha^3
- \frac{8}{9} \zeta_4 \Nf^3
+ \frac{28}{9} \zeta_3 \alpha^4
+ \frac{160}{3} \zeta_5 \Nf \alpha
\right. \nonumber \\
&& \left. ~~~
+ \frac{525}{64} \zeta_7 \alpha^2
+ \frac{560}{3} \zeta_5 \Nf^2
+ \frac{11542}{9} \zeta_4 \Nf
+ \frac{15317}{216} \zeta_3 \alpha^3
+ \frac{31595}{36} \zeta_3 \alpha^2
\right. \nonumber \\
&& \left. ~~~
+ \frac{32579}{324} \Nf \alpha^2
+ \frac{33964}{81} \zeta_3 \Nf^2
+ \frac{34204}{81} \zeta_3 \Nf \alpha
+ \frac{59275}{216} \zeta_5 \alpha
+ \frac{96979}{4374} \Nf^3
\right. \nonumber \\
&& \left. ~~~
+ \frac{3764537}{2916} \Nf \alpha
+ \frac{9369745}{432} \zeta_5
+ \frac{86284171}{324} \zeta_3
+ \frac{247516535}{2916} \Nf
- 45 \zeta_5 \alpha^3
\right. \nonumber \\
&& \left. ~~~
+ 10 \zeta_5 \Nf \alpha^2
+ 5500 \zeta_6
\right] a^4 ~+~ O(a^5)
\end{eqnarray}
for the quark mass conversion function. Finally the $SU(3)$ gauge parameter 
mapping is
\begin{eqnarray}
\left. \alpha_{\mMOMs} \right|^{SU(3)} &=& \alpha
+ \left[
\frac{10}{9} \Nf \alpha
- \frac{97}{12} \alpha
- \frac{3}{2} \alpha^2
- \frac{3}{4} \alpha^3
\right] a
\nonumber \\
&&
+ \left[
\frac{4}{3} \zeta_3 \Nf \alpha
- \frac{7143}{32} \alpha
- \frac{9}{16} \alpha^4
- \frac{5}{3} \Nf \alpha^2
- \frac{5}{3} \Nf \alpha^3
+ \frac{9}{16} \alpha^5
+ \frac{95}{16} \alpha^3
+ \frac{463}{32} \alpha^2
\right. \nonumber \\
&& \left. ~~~
+ \frac{2473}{72} \Nf \alpha
- 18 \zeta_3 \alpha^2
+ 27 \zeta_3 \alpha
\right] a^2
\nonumber \\
&&
+ \left[
\frac{15}{8} \Nf \alpha^5
- \frac{10221367}{1152} \alpha
- \frac{61105}{972} \Nf^2 \alpha
- \frac{36213}{32} \zeta_3 \alpha^2
- \frac{21763}{108} \zeta_3 \Nf \alpha
\right. \nonumber \\
&& \left. ~~~
- \frac{11417}{288} \Nf \alpha^3
- \frac{8439}{128} \alpha^4
- \frac{3049}{48} \Nf \alpha^2
- \frac{2320}{9} \zeta_5 \Nf \alpha
- \frac{1449}{32} \zeta_3 \alpha^3
\right. \nonumber \\
&& \left. ~~~
- \frac{315}{64} \zeta_5 \alpha^5
- \frac{261}{16} \alpha^5
- \frac{45}{8} \zeta_5 \alpha^4
- \frac{33}{2} \zeta_4 \Nf \alpha
- \frac{27}{64} \alpha^7
- \frac{25}{27} \Nf^2 \alpha^3
\right. \nonumber \\
&& \left. ~~~
- \frac{16}{3} \zeta_3 \Nf^2 \alpha^2
- \frac{11}{4} \zeta_3 \Nf \alpha^3
- \frac{5}{4} \Nf \alpha^4
+ \frac{4}{3} \Nf^2 \alpha^2
+ \frac{27}{16} \alpha^6
+ \frac{32}{27} \zeta_3 \Nf^2 \alpha
\right. \nonumber \\
&& \left. ~~~
+ \frac{81}{8} \zeta_4 \alpha^2
+ \frac{81}{32} \zeta_4 \alpha^3
+ \frac{117}{32} \zeta_3 \alpha^5
+ \frac{243}{32} \zeta_4 \alpha
+ \frac{1341}{32} \zeta_3 \alpha^4
+ \frac{3105}{8} \zeta_5 \alpha^2
\right. \nonumber \\
&& \left. ~~~
+ \frac{3465}{32} \zeta_5 \alpha^3
+ \frac{13941}{8} \zeta_3 \alpha
+ \frac{30835}{384} \alpha^3
+ \frac{39423}{128} \alpha^2
+ \frac{63225}{64} \zeta_5 \alpha
\right. \nonumber \\
&& \left. ~~~
+ \frac{4939411}{2592} \Nf \alpha
+ 117 \zeta_3 \Nf \alpha^2
\right] a^3
\nonumber \\
&&
+ \left[
\frac{9}{4} \zeta_4 \Nf \alpha^2
- \frac{1174691657}{2304} \alpha
- \frac{277127487}{2048} \zeta_7 \alpha
- \frac{273998929}{31104} \Nf^2 \alpha
\right. \nonumber \\
&& \left. ~~~
- \frac{70979391}{4096} \zeta_7 \alpha^2
- \frac{56971325}{2592} \zeta_3 \Nf \alpha
- \frac{54117607}{1024} \zeta_3 \alpha^2
\right. \nonumber \\
&& \left. ~~~
- \frac{18363505}{288} \zeta_5 \Nf \alpha
- \frac{13026679}{6912} \Nf \alpha^3
- \frac{12195539}{4608} \Nf \alpha^2
- \frac{2566431}{2048} \zeta_7 \alpha^3
\right. \nonumber \\
&& \left. ~~~
- \frac{1624581}{512} \zeta_3 \alpha^3
- \frac{1199043}{2048} \alpha^5
- \frac{1100549}{512} \alpha^4
- \frac{992391}{1024} \zeta_5 \alpha^4
\right. \nonumber \\
&& \left. ~~~
- \frac{851175}{1024} \zeta_6 \alpha^2
- \frac{327213}{512} \zeta_3^2 \alpha^2
- \frac{305965}{576} \zeta_4 \Nf \alpha
- \frac{226449}{512} \zeta_3^2 \alpha^3
\right. \nonumber \\
&& \left. ~~~
- \frac{216815}{96} \zeta_5 \Nf \alpha^2
- \frac{204525}{32} \zeta_6 \alpha
- \frac{137781}{1024} \zeta_7 \alpha^5
- \frac{56025}{512} \zeta_3^2 \alpha^4
\right. \nonumber \\
&& \left. ~~~
- \frac{45465}{256} \zeta_5 \alpha^5
- \frac{23247}{1024} \zeta_4 \alpha^4
- \frac{19755}{512} \zeta_5 \alpha^6
- \frac{17631}{512} \zeta_3 \alpha^6
- \frac{13041}{2048} \zeta_4 \alpha^5
\right. \nonumber \\
&& \left. ~~~
- \frac{6071}{96} \Nf \alpha^4
- \frac{5977}{216} \zeta_4 \Nf^2 \alpha
- \frac{4873}{9} \zeta_3 \Nf^2 \alpha^2
- \frac{3267}{512} \alpha^6
- \frac{1117}{8} \zeta_3 \Nf \alpha^4
\right. \nonumber \\
&& \left. ~~~
- \frac{880}{27} \zeta_5 \Nf^3 \alpha
- \frac{567}{256} \alpha^8
- \frac{351}{64} \zeta_3 \alpha^7
- \frac{175}{16} \zeta_5 \Nf \alpha^5
- \frac{128}{3} \zeta_3^2 \Nf^2 \alpha
\right. \nonumber \\
&& \left. ~~~
- \frac{97}{36} \Nf^2 \alpha^4
- \frac{40}{27} \Nf^3 \alpha^2
- \frac{15}{8} \Nf \alpha^7
- \frac{7}{2} \zeta_3 \Nf^2 \alpha^3
- \frac{5}{2} \zeta_5 \Nf^2 \alpha^3
+ \frac{9}{4} \zeta_4 \Nf^2 \alpha^2
\right. \nonumber \\
&& \left. ~~~
+ \frac{9}{4} \zeta_3^2 \Nf \alpha^4
+ \frac{23}{2} \zeta_3 \Nf \alpha^5
+ \frac{25}{12} \Nf^2 \alpha^5
+ \frac{45}{4} \zeta_3^2 \Nf \alpha^3
+ \frac{45}{8} \Nf \alpha^6
+ \frac{81}{256} \alpha^9
\right. \nonumber \\
&& \left. ~~~
+ \frac{115}{2} \zeta_5 \Nf^2 \alpha^2
+ \frac{153}{4} \zeta_4 \Nf \alpha^3
+ \frac{160}{27} \zeta_3 \Nf^3 \alpha^2
+ \frac{321}{4} \zeta_3^2 \Nf \alpha^2
+ \frac{595}{32} \zeta_5 \Nf \alpha^4
\right. \nonumber \\
&& \left. ~~~
+ \frac{945}{128} \zeta_5 \alpha^7
+ \frac{1501}{243} \zeta_3 \Nf^3 \alpha
+ \frac{1683}{8} \zeta_3^2 \Nf \alpha
+ \frac{1765}{128} \Nf \alpha^5
+ \frac{1863}{256} \zeta_3^2 \alpha^6
\right. \nonumber \\
&& \left. ~~~
+ \frac{3069}{128} \alpha^7
+ \frac{4725}{1024} \zeta_6 \alpha^5
+ \frac{10647}{4096} \zeta_7 \alpha^6
+ \frac{16775}{16} \zeta_6 \Nf \alpha
+ \frac{19575}{512} \zeta_4 \alpha^3
\right. \nonumber \\
&& \left. ~~~
+ \frac{19737}{512} \zeta_3^2 \alpha^5
+ \frac{22275}{1024} \zeta_6 \alpha^3
+ \frac{24975}{1024} \zeta_6 \alpha^4
+ \frac{26901}{32} \zeta_7 \Nf \alpha^2
\right. \nonumber \\
&& \left. ~~~
+ \frac{40279}{96} \zeta_5 \Nf \alpha^3
+ \frac{51203}{1296} \Nf^2 \alpha^3
+ \frac{87701}{576} \zeta_3 \Nf \alpha^3
+ \frac{132615}{512} \zeta_7 \alpha^4
\right. \nonumber \\
&& \left. ~~~
+ \frac{197811}{32} \zeta_3^2 \alpha
+ \frac{285995}{1728} \Nf^2 \alpha^2
+ \frac{303233}{1944} \zeta_3 \Nf^2 \alpha
+ \frac{423043}{162} \zeta_5 \Nf^2 \alpha
\right. \nonumber \\
&& \left. ~~~
+ \frac{488271}{2048} \zeta_3 \alpha^5
+ \frac{636039}{1024} \zeta_4 \alpha^2
+ \frac{1312185}{512} \alpha^3
+ \frac{2029821}{1024} \zeta_3 \alpha^4
\right. \nonumber \\
&& \left. ~~~
+ \frac{2785621}{288} \zeta_3 \Nf \alpha^2
+ \frac{2910357}{2048} \zeta_4 \alpha
+ \frac{4167895}{34992} \Nf^3 \alpha
+ \frac{4295115}{1024} \zeta_5 \alpha^3
\right. \nonumber \\
&& \left. ~~~
+ \frac{25666081}{864} \zeta_7 \Nf \alpha
+ \frac{32185629}{1024} \zeta_5 \alpha^2
+ \frac{128069033}{18432} \alpha^2
+ \frac{230899397}{2048} \zeta_3 \alpha
\right. \nonumber \\
&& \left. ~~~
+ \frac{246393489}{1024} \zeta_5 \alpha
+ \frac{541249745}{3888} \Nf \alpha
+ \zeta_4 \Nf^3 \alpha
+ 8 \zeta_3 \Nf^2 \alpha^4
\right] a^4
\nonumber \\
&& +~ O(a^5) ~.
\end{eqnarray}


\begin{thebibliography}{99}
\bibitem{1} G. 't Hooft, Nucl. Phys. {\bf B61} (1973), 455.
\bibitem{2} W.A. Bardeen, A.J. Buras, D.W. Duke \& T. Muta, Phys. Rev.
{\bf D18} (1978), 3998.
\bibitem{3} P.A. Baikov, K.G. Chetyrkin \& J.H. K\"{u}hn, Phys. Rev. Lett.
{\bf 118} (2017), 082002.
\bibitem{4} F. Herzog, B. Ruijl, T. Ueda, J.A.M. Vermaseren \& A. Vogt, JHEP
{\bf 02} (2017), 090.
\bibitem{5} T. Luthe, A. Maier, P. Marquard \& Y. Schr\"{o}der, JHEP {\bf 03} 
(2017), 020.
\bibitem{6} T. Luthe, A. Maier, P. Marquard \& Y. Schr\"{o}der, JHEP {\bf 10} 
(2017), 166.
\bibitem{7} K.G. Chetyrkin, G. Falcioni, F. Herzog \& J.A.M. Vermaseren, JHEP
{\bf 10} (2017), 179.
\bibitem{8} O. Schnetz, Phys. Rev. {\bf D97} (2018), 085018.
\bibitem{9} A.A. Vladimirov, Theor. Math. Phys. {\bf 43} (1980), 417.
\bibitem{10} K.G. Chetyrkin, A.L. Kataev \& F.V. Tkachov, Nucl. Phys. {\bf
B174} (1980), 345.
\bibitem{11} F. Herzog, Nucl. Phys. {\bf B926} (2018), 370.
\bibitem{12} K.G. Chetyrkin, G. Falcioni, F. Herzog and J.A.M. Vermaseren, 
PoS (RADCOR2017) (2018), 004. 
\bibitem{13} W. Celmaster \& R.J. Gonsalves, Phys. Rev. Lett. {\bf 42} (1979),
1435.
\bibitem{14} W. Celmaster \& R.J. Gonsalves, Phys. Rev. {\bf D20} (1979),
1420.
\bibitem{15} L. von Smekal, K. Maltman \& A. Sternbeck, Phys. Lett. {\bf B681}
(2009), 336.
\bibitem{16} J.C. Taylor, Nucl. Phys. {\bf B33} (1971), 436.
\bibitem{17} O.V. Tarasov, A.A. Vladimirov \& A.Yu. Zharkov, Phys. Lett.
{\bf B93} (1980), 429.
\bibitem{18} J.A. Gracey, J. Phys. {\bf A46} (2013), 225403; J. Phys. {\bf A48}
(2015), 119501(E).
\bibitem{19} W.E. Caswell, Phys. Rev. Lett. {\bf 33} (1974), 244.
\bibitem{20} T. Banks \& A. Zaks, Nucl. Phys. {\bf B196} (1982), 189.
\bibitem{21} R. Shrock, Phys. Rev. {\bf D89} (2014), 045019.
\bibitem{22} T.A. Ryttov, Phys. Rev. {\bf D89} (2014), 056001.
\bibitem{23} T.A. Ryttov, Phys. Rev. {\bf D90} (2014), 056007; Phys. Rev.
{\bf D91} (2015), 039906(E).
\bibitem{24} T.A. Ryttov \& R. Shrock, Phys. Rev. {\bf D94} (2016), 105015.
\bibitem{25} B. Ruijl, T. Ueda, J.A.M. Vermaseren \& A. Vogt, JHEP {\bf 06} 
(2017), 040.
\bibitem{26} A.L. Kataev \& V.S. Molokoedov, J. Phys. Conf. Ser. {\bf 938}
(2017), 012050.
\bibitem{27} A.V. Garkusha, A.L. Kataev \& V.S. Molokoedov, JHEP {\bf 02} 
(2018), 161. 
\bibitem{28} paper in preparation.
\bibitem{29} D.J. Gross \& F.J. Wilczek, Phys. Rev. Lett. {\bf 30} (1973),
1343.
\bibitem{30} H.D. Politzer, Phys. Rev. Lett. {\bf 30} (1973), 1346.
\bibitem{31} D.R.T. Jones, Nucl. Phys. {\bf B75} (1974), 531.
\bibitem{32} E. Egorian \& O.V. Tarasov, Theor. Math. Phys. {\bf 41} (1979),
863.
\bibitem{33} S.A. Larin \& J.A.M. Vermaseren, Phys. Lett. {\bf B303} (1993),
334.
\bibitem{34} T. van Ritbergen, J.A.M. Vermaseren \& S.A. Larin, Phys. Lett.
{\bf B400} (1997), 379.
\bibitem{35} M. Czakon, Nucl. Phys. {\bf B710} (2005), 485.
\bibitem{36} P.A. Baikov, K.G. Chetyrkin \& J.H. K\"{u}hn, JHEP {\bf 10} (2014),
076.
\bibitem{37} T. Luthe, A. Maier, P. Marquard \& Y. Schr\"{o}der, JHEP {\bf 01} 
(2017), 081.
\bibitem{38} P.A. Baikov, K.G. Chetyrkin \& J.H. K\"{u}hn, JHEP {\bf 04} (2017),
119.
\bibitem{39} S.G. Gorishny, S.A. Larin, L.R. Surguladze \& F.K. Tkachov,
Comput. Phys. Commun. {\bf 55} (1989), 381.
\bibitem{40} S.A. Larin, F.V. Tkachov \& J.A.M. Vermaseren, The Form version of
Mincer, NIKHEF-H-91-18.
\bibitem{41} K.G. Chetyrkin \& T. Seidensticker, Phys. Lett. {\bf B495} (2000),
74.
\bibitem{42} T. Ueda, B. Ruijl \& J.A.M. Vermaseren, PoS (LL2016) (2016), 070.
\bibitem{43} T. Ueda, B. Ruijl \& J.A.M. Vermaseren, Comput. Phys. Commun.
{\bf 253} (2020), 107198.
\bibitem{44} J.A.M. Vermaseren, math-ph/0010025.
\bibitem{45} M. Tentyukov \& J.A.M. Vermaseren, Comput. Phys. Commun. {\bf 181}
(2010), 1419.
\bibitem{46} J.A. Gracey, arXiv:2210.12420 [hep-ph].
\bibitem{47} O. Nachtmann \& W. Wetzel, Nucl. Phys. {\bf B187} (1981), 333.
\bibitem{48} R. Tarrach, Nucl. Phys. {\bf B183} (1981), 384.
\bibitem{49} O.V. Tarasov, Phys. Part. Nucl. Lett. {\bf 17} (2020), 109.
\bibitem{50} K.G. Chetyrkin, Phys. Lett. {\bf B404} (1997), 161.
\bibitem{51} J.A.M. Vermaseren, S.A. Larin \& T. van Ritbergen, Phys. Lett.
{\bf B405} (1997), 327.
\bibitem{52} O. Bergner, P. Giudice, I. Montvay, G. M\"{u}nster, S. Piemonte \&
P. Scior, PoS (LATTICE2018) (2018), 191.
\bibitem{53} V. Afferrante, A. Maas, R. Sondenheimer \& P. T\"{o}rek,
SciPost Phys. {\bf 10} (2021), 062.
\end{thebibliography}
\end{document}